\documentclass[sigconf]{acmart}
\AtBeginDocument{%
  }

\copyrightyear{2025}
\acmYear{2025}
\setcopyright{cc}
\setcctype{by}
\acmConference[CHI '25]{CHI Conference on Human Factors in Computing Systems}{April 26-May 1, 2025}{Yokohama, Japan}
\acmBooktitle{CHI Conference on Human Factors in Computing Systems (CHI '25), April 26-May 1, 2025, Yokohama, Japan}\acmDOI{10.1145/3706598. 3713599}
\acmISBN{979-8-4007-1394-1/25/ 04}

\usepackage{colortbl} %
\usepackage{multirow} 
\usepackage{graphicx} 
\usepackage{balance}
\usepackage{stfloats} 

\definecolor{darkgray}{HTML}{333333} 
\definecolor{midgray}{HTML}{444444}

\begin{document}
\title{Layered Interactions: Exploring Non-Intrusive Digital Craftsmanship Design Through Lacquer Art Interfaces}

\author{Yan Dong}
\authornote{Both authors contributed equally to this research.}
\orcid{0009-0008-5989-4201}
\affiliation{
 \institution{Department of Information Art \& Design, Academy of Arts \& Design, Tsinghua University, Beijing, China}
 \city{}
 \country{}
 }
\email{dongyan0111@outlook.com}

\author{Hanjie Yu}
\authornotemark[1]
\orcid{0009-0000-9085-5122}
\affiliation{
 \institution{Department of Information Art \& Design, Academy of Arts \& Design, Tsinghua University, Beijing, China}
 \city{}
 \country{}
 }
\email{yuhanjie.yhj@gmail.com}

\author{Yanran Chen}
\orcid{0000-0002-0475-2432}
\affiliation{%
 \institution{Department of Information Art \& Design, Academy of Arts \& Design, Tsinghua University, Beijing, China}
  \city{}
  \country{}
 }
\email{chenyr0909@gmail.com}

\author{Zipeng Zhang}
\orcid{0009-0005-2047-5278}
\affiliation{%
 \institution{Pervasive Human Computer Interaction Laboratory, Tsinghua University, Beijing, China}
  \city{}
  \country{}
 }
\email{guanchaxiao93@gmail.com}

\author{Qiong Wu}
\authornote{Denotes the corresponding author.}
\orcid{0000-0002-0304-5330}
\affiliation{%
 \institution{Department of Information Art \& Design, Academy of Arts \& Design, Tsinghua University, Beijing, China}
 \city{}
 \country{}
 }
\email{qiong-wu@mail.tsinghua.edu.cn}


\begin{abstract}
    Integrating technology with the distinctive characteristics of craftsmanship has become a key issue in the field of digital craftsmanship. This paper introduces Layered Interactions, a design approach that seamlessly merges Human-Computer Interaction (HCI) technologies with traditional lacquerware craftsmanship. By leveraging the multi-layer structure and material properties of lacquerware, we embed interactive circuits and integrate programmable hardware within the layers, creating tangible interface that support diverse interactions. This method enhances the adaptability and practicality of traditional crafts in modern digital contexts. Through the development of a lacquerware toolkit, along with user experiments and semi-structured interviews, we demonstrate that this approach not only makes technology more accessible to traditional artisans but also enhances the materiality and emotional qualities of interactive interfaces. Additionally, it fosters mutual learning and collaboration between artisans and technologists. Our research introduces a cross-disciplinary perspective to the HCI community, broadening the material and design possibilities for interactive interfaces.  
\end{abstract}



\begin{teaserfigure}
  \includegraphics[width=\textwidth]{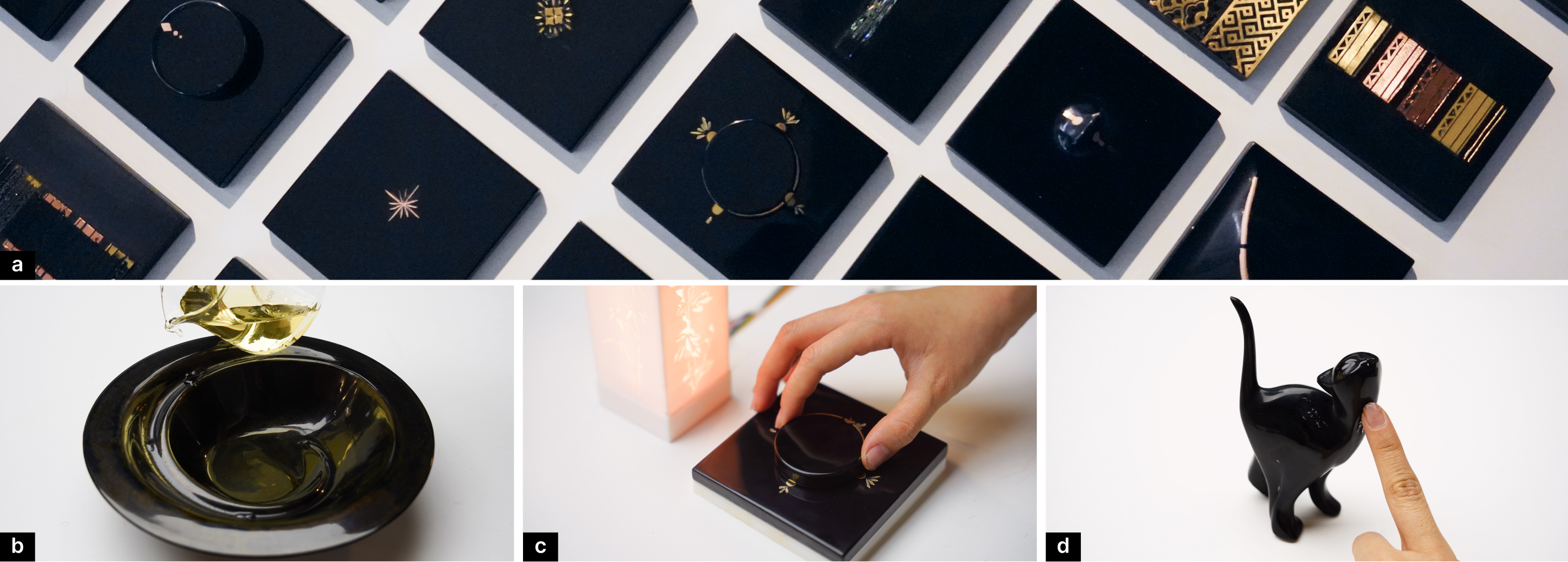}
  \caption{The integration of HCI with traditional lacquerware has resulted in the design of several digital lacquerware products, including: (a) Lacquer art interface; (b) Musical tableware; (c) Artistically designed input unit; (d) Interactive cat figurine.}
  \Description{This figure presents a wide range of applications for interactive lacquer interfaces, demonstrating how traditional lacquer craftsmanship can seamlessly integrate with interactive technology. These interfaces feature smooth, visually refined surfaces that support various interaction modalities, offering new possibilities for physical-digital experiences across functional, decorative, and playful applications.
  (a) Lacquerware interface – A collection of lacquer interfaces showcasing different craftsmanship techniques and embedded interactive elements. (b) Interactive Lacquer Tableware – A lacquer bowl embedded with a temperature sensor, responding dynamically to liquid temperature changes. (c) Lacquer interface application (Knob) – A Jinyinpingtuo-based lacquer interface functioning as a rotary controller to adjust the brightness of a lamp. (d) Interactive cat figurine – A lacquer figurine embedded with a pressure sensor, enabling playful interactions through touch-based input.}
  \label{fig:teaser}
\end{teaserfigure}


\maketitle
\section{Introduction}
More and more Human-Computer Interaction (HCI) researchers recognize the need for integrating digital technology with traditional crafts \cite{10.1145/3613904.3642361, 10.1145/3613904.3642205, 10.1145/1640233.1640264, 10.1145/3569009.3576187}, numerous exploratory efforts have emerged within design practice, driving innovation in the creation and production of digital craftsmanship  \cite{10.1145/3613904.3642361, 10.1145/3613904.3642684, 10.1145/3290607.3299010, 10.1145/3569009.3576187}.  Despite the increasingly evident trend of merging digital technology with traditional craft, digital craftsmanship remains a contested concept \cite{10.1145/3569009.3572746}. This paper aims to explore design methods that organically blend the unique qualities of both technology and craftsmanship through practical experimentation.

The dual origins of digital craftsmanship add to the complexity of integration. It involves not only the convergence of craft and technology but also the interplay of materials, tools, production processes, and individual skills. Peizhong Gao et al. proposed that the degree of integration determines both the process and the product, with the boundaries of this "degree" depending on the designer's intent, skill, and evaluation \cite{10.1145/3569009.3572746}. Justin Marshall further emphasized a "digital craft spirit," which focuses on fidelity rather than precision, and promotes coexistence with handcraft rather than replacing it \cite{10.1145/3623509.3635316}. Despite general consensus that collaborative creation could potentially break down communication barriers between digital manufacturing and traditional craftsmanship \cite{10.1145/2702123.2702362, 10.1145/3313831.3376820}, questions remain about how to effectively integrate technology and craft to fully unlock the potential of digital craftsmanship.

Since the late 18th century, the tension between these two domains has been apparent, reflecting the challenge of achieving seamless integration \cite{10.1145/3613904.3642361, 10.1145/1640233.1640264, 10.1145/3290607.3299010}. Digital technologies prioritize efficiency and rapid production, whereas craft values meticulous, time-intensive processes, creating a fundamental conflict between speed and quality \cite{10.1145/3569009.3572746, kreitzberg2019careers}. This tension also extends to the roles of artisans and technologists, with digital technologies sometimes presenting barriers for artisans unfamiliar with new paradigms \cite{10.1145/3613904.3642361}. Reintegrating digital workflows into non-industrial environments introduces both technical and conceptual challenges that often clash with the values of independent designers and artisans \cite{10.1145/3623509.3635316}. Handmade artifacts emphasize uniqueness, while mass-produced objects tend to be associated with computation, automation, and batch production, which can feel less empathetic and natural \cite{10.1145/3623509.3635316, 10.1145/3613904.3642205}. Moreover, the exponential growth of digital technologies and ubiquitous computing is altering the functionality of physical objects, which may, in turn, change our emotional connections to them \cite{10.1145/3613904.3642684}. Designers are expected to adopt a more integrated perspective that blends cultural context with technical tools, empowering human craftsmanship rather than confining it within the limitations of the digital realm \cite{10.1145/2702123.2702362}. Through practice, we aim to find a middle ground to reconcile these tensions.

This paper focuses on lacquer art, a craft with limited exploration in the digital craft domain. Inspired by its multi-layered structure, distinctive material properties, and diverse artistic techniques, we develop innovative design practices. Our research advocates for HCI researchers to explore non-invasive digital craft practices that preserve traditional lacquer art methods, processes, and materials. We aim to redefine digital tools as extensions of the artist's creative process, enabling richer expression through technology.

Our contributions include:

\begin{enumerate}
    \item We propose a new method for seamlessly integrating digital technology with traditional craft. Through integrating interactive circuits within the multi-layered process of lacquer art, we create tangible user interfaces (TUI) that support multiple lacquer techniques and interaction modes. Leveraging lacquer's unique tactile and aesthetic qualities, along with its wide range of applications, we enhance the adaptability and practicality of traditional craft in modern digital environments. This also enriches the authenticity and emotional connection of artifacts, allowing digital interactions to blend more naturally into everyday life.    \item An intelligent lacquer art toolkit. By breaking down craft and technical modules, we map the key steps in craft processes to core technical concepts, fostering mutual understanding across disciplines and facilitating bi-directional learning and collaboration between artisans and technologists. Additionally, the toolkit enables traditional craftsmen to more easily engage with digital technologies, expanding their creative possibilities and bringing the craft community into HCI research and practice.
\end{enumerate}

\section{Related Works}
\subsection{Craft-Centered Digital Craft}
The initial practices of digital craftsmanship were predominantly technology-centric, integrating digital resources to improve the efficiency, expressive methods \cite{10.1145/3623509.3633359, 10.1145/3623509.3633353}, and production capacity of craft processes \cite{kreitzberg2019careers, 10.1145/3313831.3376820}. This allowed artisans to focus more on creativity while reducing labor-intensive tasks \cite{10.1145/3569009.3572746, kreitzberg2019careers}. For instance, technologies such as Computer-Aided Design (CAD), Computer-Aided Manufacturing (CAM), 3D printers, laser cutters, and Computer Numerical Control (CNC) machines have been widely used in the creation of works with complex structures \cite{kreitzberg2019careers}, precise shapes \cite{10.1145/3623509.3633363}, or unique surface textures  \cite{10.1145/3411764.3445600}. Software algorithms have also been employed to assist in the design process of craft objects \cite{mueller2012interactive, 10.1145/3290605.3300554}. Even complex 3D-printed components have been incorporated into crafts \cite{10.1145/2702123.2702362, 10.1145/3544548.3581298}. However, digital automation also have challenges. The automated functionalities of digital tools may cause creators to over-rely on predefined models and algorithms, thus weakening personalized expression \cite{10.1145/3613904.3642361, 10.1145/3196709.3196750}. For example, many slicers automatically generate toolpaths for users \cite{10.1145/3544548.3580745, 10.1145/3322276.3323694}, and CNC machines are said to be "operable by a monkey" \cite{noble2013forces}. This indicates that digital manufacturing-centric practices \cite{10.1145/3613904.3642361} can overlook the creativity and emotional expression of artisans in the creation process\cite{noble2013forces}, making the works overly "perfect" and diminishing the uniqueness, irregularities, and rich texture of handmade crafts.

Some researchers have reflected on this technology-dominant practice \cite{10.1145/3196709.3196750}, proposing craft-centered design approaches \cite{10.1145/3613904.3642361, 10.1145/2675133.2675291}. Ilan E. Moyer and others have modified CNC systems based on the needs of artisans, bringing digital manufacturing back to the core of skilled craft practice \cite{10.1145/3613904.3642361}. The Digital Pottery Wheel (DPW), while innovative, remains constrained to vessel-like forms, potentially limiting creative exploration. Devon Frost designed the SketchPath system, which allows artists to hand-draw toolpaths for creating clay 3D-printed forms \cite{10.1145/3613904.3642205}. However, relying on digital interfaces may reduce the artist's intuitive interaction with materials and physical perception \cite{10.1145/3623509.3633351}, with the risk of losing information about the physical characteristics of artifacts. This approach may also limit the creative and making processes of artisans due to the constraints imposed by digital manufacturing tools. Therefore, the main challenge in integrate digital technologies in a way that enhances the expressiveness of craftsmanship without diminishing the artisan’s creativity. Our approach starts with the intrinsic processes and characteristics of craftsmanship, restoring creative leadership to artisans while preserving the uniqueness of traditional culture.

\subsection{Materiality of Craft-Enhanced HCI Technologies as Tangible Bits Interfaces}
Early concepts such as Tangible Bits \cite{ishii1997tangible, 10.1007/3-540-69706-3_4, 2_3_4}aimed to transition from virtual interfaces to the physical world, but the materiality of current systems remains weak \cite{10.1145/3623509.3633359, 10.1145/3623509.3633351}. The incorporation of craft brings a new dimension to the materiality of tangible bits interfaces \cite{10.1145/3569009.3572744, nordmoen2022making}, diversifying the forms and materials of interaction between the digital and physical worlds \cite{10.1145/3544548.3580836, 2_3_10}, making them more embodied and concrete \cite{10.1145/2702123.2702362, ishii1997tangible, 10.1145/3544548.3580836, 2_3_12, 10.1145/3613904.3642010}, while also enhancing their aesthetic qualities and cultural significance \cite{7317459, 10.1145/2908805.2913018, 10.1145/3623509.3633353}. For example, Zheng et al. explored glazed ceramics as a platform for interactive interfaces,  highlighting how everyday artifacts and craft practices can enhance the diversity and expressiveness of ubiquitous computing systems \cite{10.1145/3544548.3580836}. Knitting Interaction Spaces demonstrated how digital information can be materialized through textile fabrication machines \cite{10.1145/3623509.3633359}, enhancing the realism of virtual materials and the expressiveness of digital information in physical forms. Sensing Kirigami\cite{zheng2019sensing} further investigated the material properties of carbon-coated paper and its applicability in tangible interactions, showing how cut-paper materials and their structures provide new functionalities for tangible interfaces. Vasiliki Tsaknaki et al. \cite{tsaknaki2017articulating} emphasized how the unique physical properties of silversmithing, such as malleability and conductivity, inspire interaction design.

Moreover, digital craftsmanship helps create more effective cultural interfaces, transforming intangible culture into tangible objects through craft. Jennifer Jacobs et al. combined digital tools with traditional ostrich eggshell jewelry-making in South African communities,  reflecting on the potential homogenization of cultural expression brought about by the proliferation of digital technologies. They proposed that craft can serve as a means of cross-cultural communication \cite{10.1145/2702123.2702362}. Jayne Wallace et al. embedded RFID readers in handmade artifacts to trigger memories for dementia patients, demonstrating how digital technologies can amplify the narrative value of crafted objects. This allows users to resonate emotionally with the stories behind them, thus making the artifacts more enduring \cite{10.1145/2470654.2481363}. These studies have made significant contributions to the integration of HCI and traditional crafts, affirming the role of craft in enhancing and advancing HCI, which provides valuable insights for our research. Our study further explores how craft can enhance the materiality of HCI technologies through lacquer art.

\subsection{Collaborative Models Across Disciplines in Digital Craft}
Interdisciplinary collaboration is essential for improving the quality of digital craftsmanship \cite{10.1145/3341163.3346943}. Engineers and artists collaborate by combining their respective expertise to address multi-dimensional challenges such as technical complexity, material selection, and creative expression \cite{10.5555/3017447.3017560}. Despite the potential benefits, acquiring the necessary digital skills remains a significant hurdle for traditional artisans. Acquiring digital knowledge, tools, and operational skills often requires considerable time and resources, which may affect their creative flexibility and spontaneity. While artisans appreciate tools modified by HCI practitioners, they often seek greater creative autonomy. \cite{10.1145/3613904.3642361}.
Alyshia Bustos  et al.  introduced interactive murals as a new space for collaborative STEAM learning, helping students with no prior experience in electronics or programming to gradually acquire essential skills through the design and construction of these murals \cite{bustos2024interactive}. Similarly, Camila Friedman-Gerlicz \cite{friedman2024weaveslicer}, Devendorf, and others \cite{10.1145/3313831.3376820} clarified that artist residency programs facilitate collaboration between artists and HCI researchers, with tool design driven by the artists' practical needs. This problem-oriented approach underscores the advantages of integrating craft methods into engineering research.

Successful collaboration requires overcoming differences in terminology, methods, and expectations. Research highlights the importance of shared understanding and goal alignment \cite{10.5555/3017447.3017560}. Mei Zhang et al. demonstrated that translating artisans’ tacit knowledge into shared language facilitates collaboration, with boundary objects like toolkits enabling designers to explore materials without deep technical expertise \cite{10.1145/3532106.3533535}. Kristina Andersen et al. developed structured methods to document interdisciplinary collaboration, establishing a knowledge-sharing framework for products, systems, and service \cite{10.1145/3341163.3346943}. Similarly, Toka et al. \cite{toka2024practice} proposed a practice-oriented software development model that aligns computational tools with specific craft workflows. In contrast to previous research, our study is also practice-oriented but takes a step further by introducing a modular toolkit and a "Craft-Tech Alignment Manual" to refine and standardize the interdisciplinary collaboration process. Our approach not only bridges the cognitive gap between artists or craftsmen and technology developers but also simplifies the understanding of technical complexities in each other's fields through tools and manuals, while encouraging technical experts to reconsider the core value of materials and craft. Our research emphasizes workflow alignment and enhancing universality and scalability through standardized collaboration models, making it applicable to a broader range of scenarios.

\begin{figure*}[t] 
  \centering
  \includegraphics[width=\textwidth]{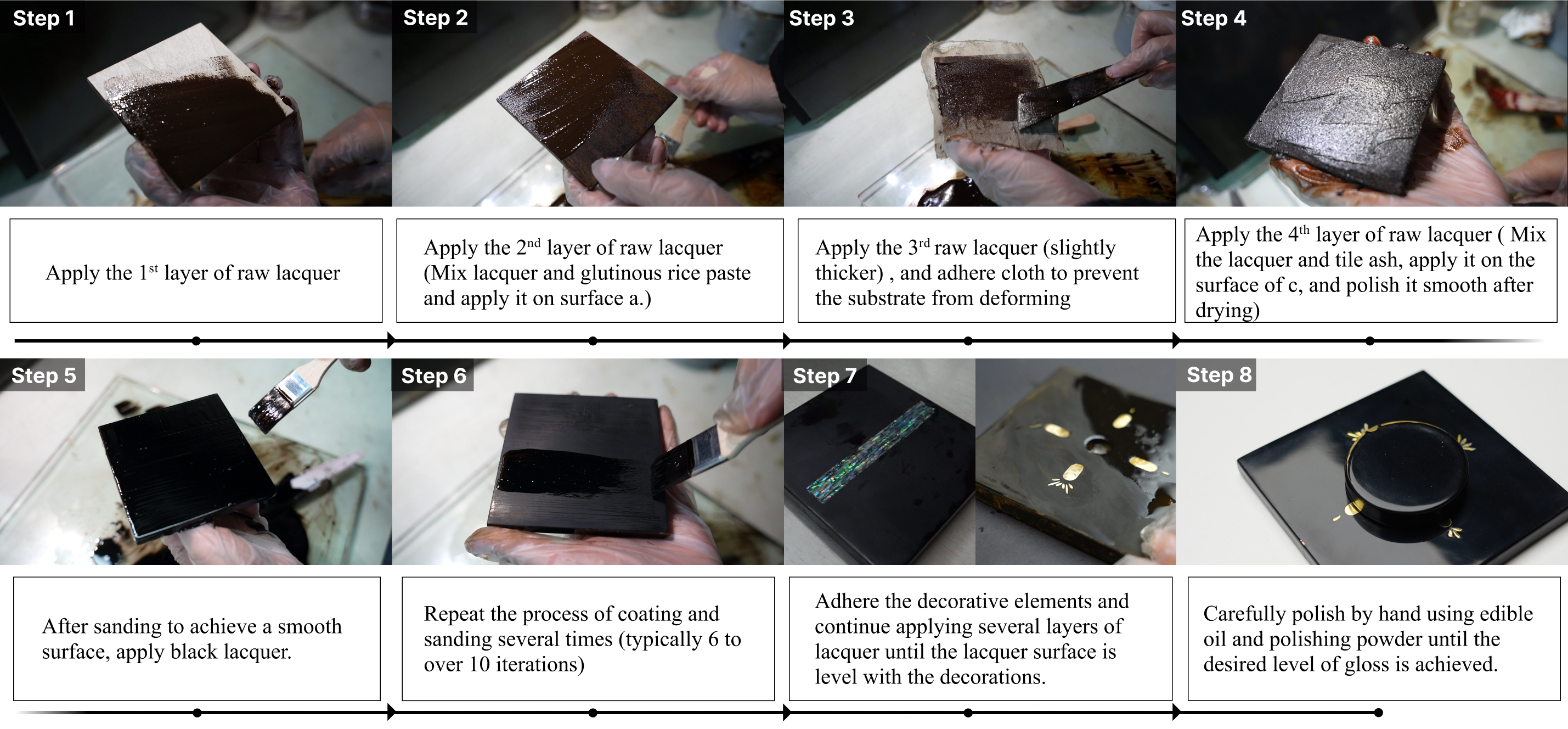}
  \caption{Traditional lacquer art production process.}
  \label{fig:lacquerprocess}
  \Description{
    A diagram showing 8 steps of traditional lacquer art production: 
    (1) Apply the 1st layer of raw lacquer. (2) Apply the 2nd layer of raw lacquer (Mix lacquer and glutinous rice paste and apply it on surface a.). (3) Apply the 3rd raw lacquer (slightly thicker) , and adhere cloth to prevent the substrate from deforming. (4) Apply the 4th layer of raw lacquer ( Mix the lacquer and tile ash, apply it on the surface of c, and polish it smooth after drying) (5) After sanding to achieve a smooth surface, apply black lacquer. (6) Repeat the process of coating and sanding several times (typically 6 to over 10 iterations) (7) Adhere the decorative elements and continue applying several layers of lacquer until the lacquer surface is level with the decorations. (8) Carefully polish by hand using edible oil and polishing powder until the desired level of gloss is achieved.}
\end{figure*}

\subsection{Lacquerware as a Design Space in Digital Craftsmanship}
Lacquer, a natural coating material, is often applied to objects due to its unique aesthetics and decorative properties \cite{10.1145/3569009.3572746}. It has a long history in China, Japan, Korea, and Southeast Asia, and is widely used in daily objects such as utensils, musical instruments, and furniture \cite{2_1_3}. However, with the advent of modern industrialization and the influx of foreign cultures, the utility value of lacquer art has gradually become blurred. In the digital age, consumers are more focused on smart and interactive products, and traditional lacquerware lacks such features. Additionally, compared to handmade products, lacquerware is more complex and costly to produce. Furthermore, lacquer art is a highly personalized form of tacit knowledge, often passed down through workshop or familial apprenticeship models, limiting its accessibility. The allergenic risks associated with raw lacquer also increase the communication barriers \cite{2_2_14}. Current practitioners mainly include master craftsmen, restoration experts, and craft companies, with limited knowledge of digital technologies and few interdisciplinary collaborations.

While digital craftsmanship has been widely explored in fields like ceramics and textiles, there has been relatively little research on the digital application of lacquerware. Researchers have employed various digital technologies to enhance traditional lacquer techniques. Laser cutting has been used to address slotting in the maki-e technique \cite{2_2_Kenji1}, and spray coating processes have significantly reduced lacquering time  \cite{2_2_young}. Additionally, 3D printing has simplified the production of lacquerware substrates \cite{yan2024applications}. Studies on lacquer’s chemical properties have explored strengthening surface hardness and applications in architecture and industry \cite{2_1_10}. Despite these digital initiatives, the disconnect between lacquer techniques and contemporary life remains unresolved. Subsequent researchers have attempted to integrate interactive circuits with lacquer. Naoya Koizumi explored a new method of designing lacquer surfaces using UV projection \cite{2_2_13}, while Keita Saito integrated NFC antenna patterns into lacquered vessels \cite{10.1145/3374920.3374952}. However, these studies have not fully captured the essence of traditional lacquer art. In our research, we combine the widespread application of lacquer with ubiquitous computing, utilizing its multilayered structure to create lacquer interfaces, ensuring the preservation and innovation of traditional crafts through daily use.

\section{Lacquer Art Interfaces}
\subsection{Decomposing the chracteristics and production process of lacquerware}
From an HCI perspective, making lacquerware involves deep interaction between the artisan and the material. Through repeated processes of brushing, sanding, and decorating, artisans engage in a “dialogue” with the material, inspiring interaction design. Based on this understanding, we advocate respecting the craft process of lacquerware in technical design, rather than merely "grafting" existing technologies onto traditional crafts. Through this realization, we have discovered that the intricate production process not only ensures the unique tactile qualities of lacquer surfaces, but also provides a feasible avenue for embedding interactive elements into the multilayered structure. Electronic components can be seamlessly integrated within these layers, leveraging the material’s unique advantages. Specifically, this approach operates in the following ways:

\leftskip 10pt
\emph{Advantage 1: Integration of Aesthetics and Functionality.} The thin lacquer surface allows the integration of sensors and other components at various layers and depths, enabling complex interfaces without significantly altering the appearance or tactility of the lacquerware.

\emph{Advantage 2: Unique Tactile Interface\cite{2_1_6}.} The smooth, skin-like texture of lacquer is ideal for touch-sensitive interfaces, capturing user touch and swipe gestures to provide a natural experience.

\emph{Advantage 3: Protection of Electronic Components.} Lacquer’s self-leveling properties tightly encase electronic components, providing natural insulation and protection from interference without additional layers \cite{HanadaNobukokoro2022}. At room temperature, lacquer has strong antibacterial properties \cite{2_1_8} and hardness \cite{2_1_10}, offering corrosion protection for materials such as metals and ceramics.

\emph{Advantage 4: Excellent adhesive}\cite{2_1_7}. Lacquer can bond with various materials, including wood, metal, leather, ceramics, bone, paper, and fabric. With traditional techniques (such as "Tuotai") to create complex shapes. Lacquer's ability to adhere to various shapes and materials makes it well-suited for ubiquitous computing applications.

\leftskip 0pt    

Traditional lacquer art is known for its thin and smooth material properties, skin-like tactile qualities, and multilayered brushing technique \cite{2_1_6}. The construction of high-quality lacquerware involves multiple intricate steps (Figure \ref{fig:lacquerprocess}). Generally, a shaded room is required for drying lacquerware. In the shaded room, humidity is maintained at around 65\%–85\% and temperature at 20–25°C. Each layer of lacquer must be air-dried in the shaded room before applying the next layer. Even the most basic lacquering technique (applying lacquer solely for surface decoration on a pre-made substrate) involves repetitive lacquering, curing, and sanding (using sandpaper ranging from 180 to 5000 grit) until the surface is mirror-smooth, ready for decoration. A thin lacquer surface can exceed 10 layers, while complex carved lacquer pieces may require 200 to 300 layers. For instance, in our work, the lacquerware crafting process includes the following eight steps:

\leftskip 12pt
\emph{Step 1: Anti-Corrosion.} Sanded the wood base first. Prepare a raw lacquer solution by mixing raw lacquer and camphor oil (Fuzhou Dongchang Raw Paint Co., Ltd) in a 1.2:1 ratio (by weight), then evenly brush it onto the base. Allow the lacquered wood to air dry in a shaded room. Exhale onto the lacquer surface; if the moisture evaporates quickly, the surface is dry and the next step can begin. Once dry, sand the surface with 320-grit sandpaper to remove particles. Reapply the lacquer using the same method to further seal the wood and enhance its anti-corrosion properties. After drying, sand the surface again with 320-grit sandpaper for a smoother finish.

\emph{Step 2-4: Shaping.} The purpose of this step is to prevent the wood base from cracking and to ensure its durability. Sticky rice powder and hot water are mixed in a 1:4 ratio, stirred into a paste, and allowed to cool. This paste is then blended with the raw lacquer solution in a 3:1 ratio. Coarse hemp fabric is cut to fit the base. The resulting mixture is evenly applied to the coarse hemp fabric and wood base, ensuring smooth adhesion of the cloth to the wood. The piece is left to air dry in a shaded room. Once dry, the surface is sanded using 320-grit sandpaper. The raw lacquer solution is mixed with tile ash power  in a 2:1 ratio. The mixture is spread evenly over the surface with a scraper, enhancing the density to achieve better gloss and form for the top lacquer layers. The piece is left to air dry in a shaded room. Once dried, the surface is sanded with 480-grit sandpaper.

\emph{Steps 5–6: Suxiu} \footnote{A traditional Chinese monochrome lacquerware technique characterized by simplicity, with no patterns and the use of a single lacquer color for decoration.}. Decorated solely with multiple layers of single-color lacquer. Black lacquer (Fuzhou Dongchang Raw Paint Co., Ltd) is mixed with camphor oil in a 1:1.2 ratio. The mixture is applied in a extremely thin layer to the surface. Once dried, the surface is polished using 1000–3000 grit sandpaper. This process of application, drying, and sanding is repeated until the surface becomes smooth and even.

\emph{Step 7: Decorative Patterns.} Appropriate techniques and materials are selected for surface decoration, typically requiring at least 2–3 layers of lacquer to achieve intricate details. Common techniques include Suxiu, Miaoqi\footnote{A technique of painting patterns with colored lacquers on a smooth lacquer base.}, Qiangjin\footnote{A technique in which lines or fine points are carved into a lacquered surface, followed by filling the carvings with gold lacquer or applying gold leaf, creating golden patterns.}, Jinyinpingtuo\footnote{ A technique in which gold and silver are melted into foil sheets, cut into patterns, and applied to the surface of lacquerware. After several layers of lacquer are applied and dried, the surface is polished to reveal the gold and silver beneath.}, Luodian/Eggshell\footnote{A decorative technique using thin eggshell or mother-of-pearl slices to create images, geometric patterns, or text, which are inlaid onto the surface of the object.}, and Xipi\footnote{ A type of lacquerware in which colored lacquer is applied in layers to a lacquer base, then polished to reveal beautiful textures, a technique with a long historical tradition.}, et al. \cite{webb2000lacquer}\cite{2022Xiushilu}.

\emph{Step 8: Luster}. Finally, the lacquerware is polished to achieve a mirror-like shine. Using vegetable oil and fine ash, the surface is repeatedly rubbed and buffed by hand until it becomes bright and reflective like a mirror.

\leftskip 0pt

\begin{figure}[H]
\vspace{-4pt}
\centering
\includegraphics[width=\linewidth]{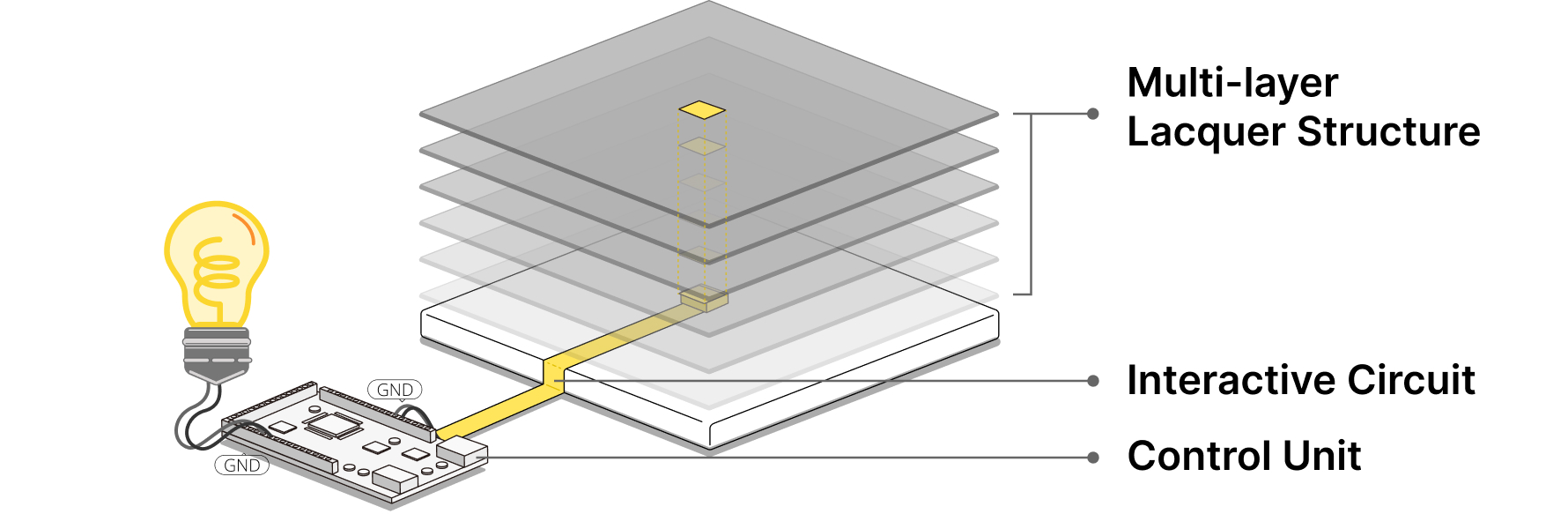}
\vspace{-18pt}
\caption{Mechanism and structure of lacquer art interfaces.}
\vspace{-10pt}
\label{fig:basicmechanism}
\Description{Diagram illustrating the mechanism and structure of lacquer art interfaces. The system consists of three main components: an interactive circuit, a multi-layer lacquer structure, and a control unit.}
\end{figure}

\subsection{Mechanism and Structure}
Based on the existing multi-layer lacquer application process and the integration advantages of lacquer and interactive circuits, we have designed a specific mechanism that combines embedded interactive circuits with traditional lacquer art techniques (Figure \ref{fig:basicmechanism}), seamlessly integrating digital technology into lacquer art, creating functional and aesthetically pleasing Lacquer Art Interfaces. This design consists of three key components:

\begin{enumerate}
\item {\emph{Interactive Circuit}}: The interactive circuit is composed of copper foil conductors and sensors embedded within the lacquer layers, transmitting user inputs to the control unit. The copper foil (thickness < 0.1 mm) can be fully covered by 2-3 layers of lacquer without affecting the appearance of the artwork. Its flexibility allows it to adapt to various surface curvatures, preventing deformation over time. Sensors, soldered to the foil, capture interactions and relay signals to the control unit.
\item {\emph{Multi-layer Lacquer Structure}}: The lacquer layers provide insulation, protecting the circuits and sensors. At the same time, lacquer layers of varying depths can accommodate multiple interactive components, ensuring that circuits between different layers do not interfere with each other, while preserving the traditional aesthetic of lacquerware.
\item {\emph{Control Unit}}: Powered by controllers like Arduino, the control unit processes signals and manages interactions. The code within the control unit can be modified to suit various application scenarios, offering scalability and flexibility to the system.
\end{enumerate}

\begin{figure*}[t]
\centering
\includegraphics[width=0.99\textwidth]{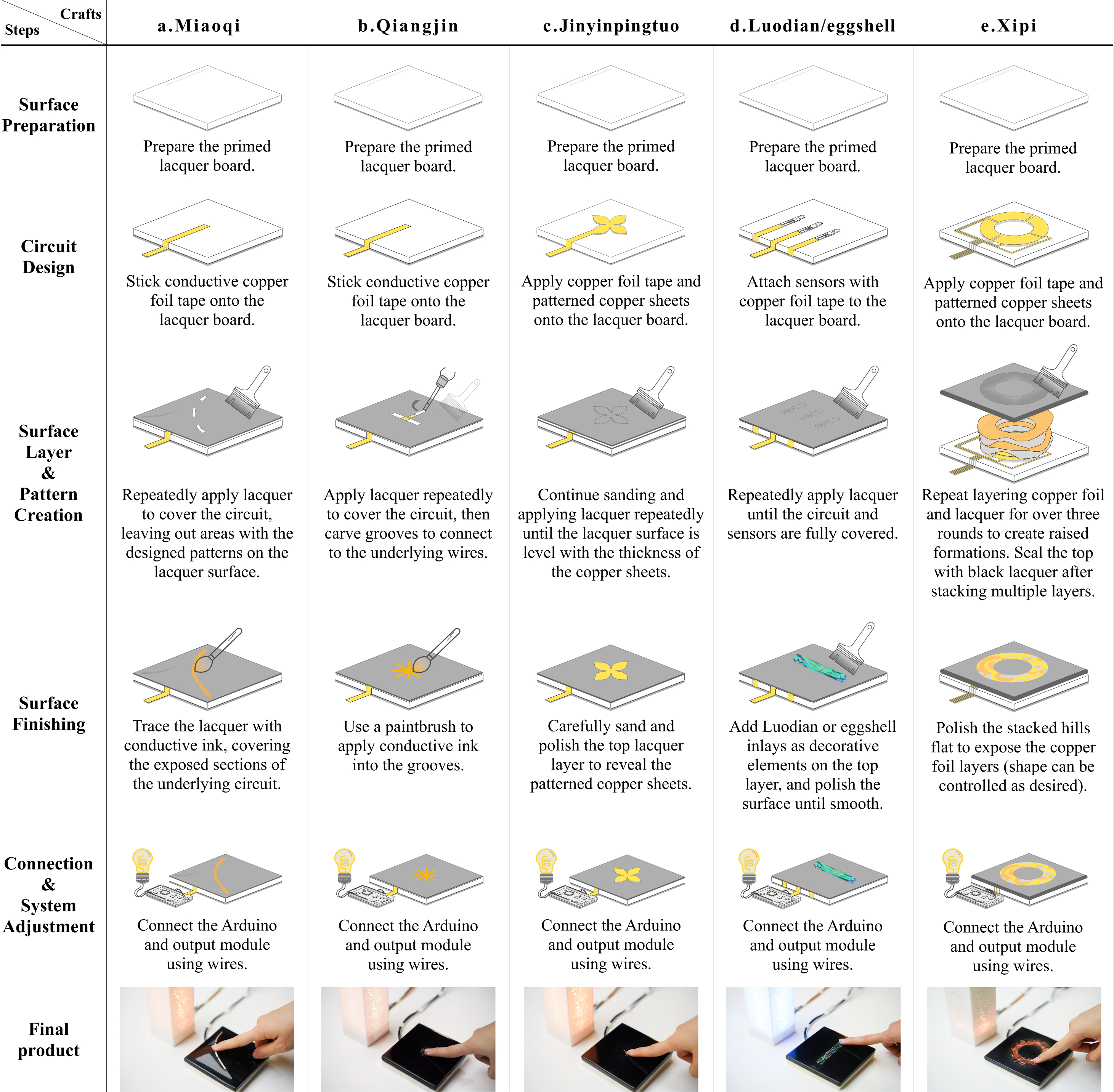}
\caption{The seamless integration of embedded interactive circuits within traditional craft processes. (a) Miaoqi: electronic ink is used to draw patterns on the base lacquer surface. The patterns and circuit contact points must remain connected to ensure functionality; (b) Qiangjin: a groove is carved into the lacquer surface using a fine knife, into which electronic ink is injected. The patterns and circuit contact points need to stay connected for the circuit to function; (c) Jinyinpingtuo: metal patterns, such as gold, silver, or copper, are embedded into the base lacquer surface. The patterns and circuit contact points must remain connected. The metal patterns are then completely covered by lacquer layers, and carefully sanded to reveal them; (d) Luodian/Eggshell: sensors are attached to the base lacquer surface, with the option to expose or conceal the interactive circuits depending on the desired effect circuits can be hidden beneath the lacquer layers. Finally, shell inlays are embedded into the surface; (e) Xipi: a textured surface is created on the base lacquer by embedding metal materials between multiple lacquer layers. The metal and circuit contact points must stay connected. Polishing reveals different layers of metal, forming a conductive path when the surface is touched.}
\label{figure4}
\Description{This figure presents a comparative workflow of Layered Interaction across five traditional lacquerware techniques: Miaoqi, Qiangjin, Jinyinpingtuo, Luodian/Eggshell, and Xipi. The process is structured into six stages: (1) Surface Preparation – A primed lacquer board is prepared as the base for all methods. (2) Circuit Design – Conductive copper foil tapes and patterned copper sheets are applied to integrate interactive circuits into the lacquerware. (3) Surface Layer and Pattern Creation – Multiple layers of lacquer are applied to embed the circuit within the surface while maintaining aesthetic integrity. (4) Surface Finishing – Additional decorative steps such as conductive ink tracing (Miaoqi), gold-filled grooves (Qiangjin), polished copper sheets (Jinyinpingtuo), decorative inlays (Luodian/Eggshell), and sculpted surface formations (Xipi) refine the visual and tactile quality. (5) Connection and System Adjustment – An Arduino module and output device are connected to the embedded circuit for interactivity. (6) Final Product – The completed lacquer interface integrates aesthetic and interactive properties, as demonstrated in the final application images.}
\end{figure*}
\clearpage

\begin{figure*}[ht]
\centering
\includegraphics[width=\textwidth]{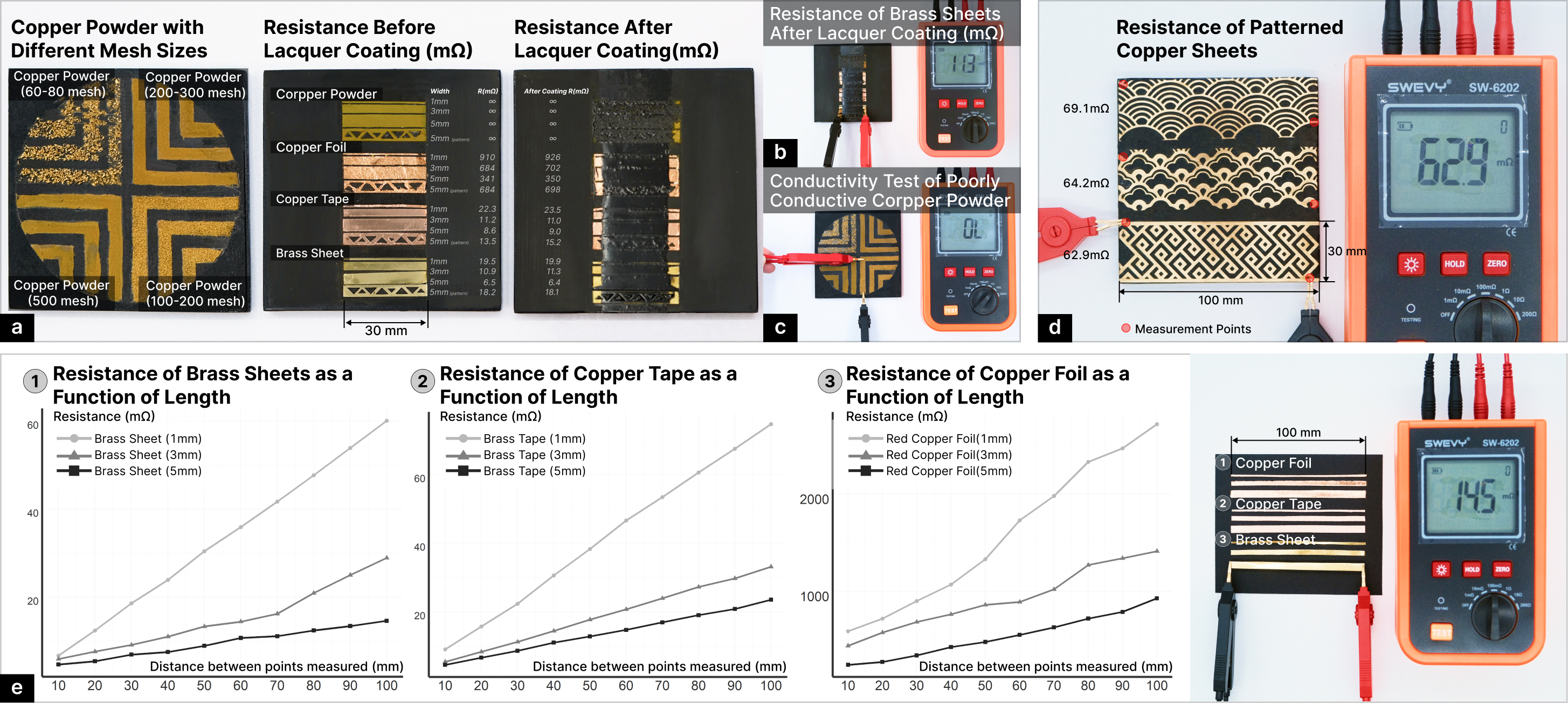}
\vspace{-0.5cm}
\caption{Performance verification. (a) Resistance of metal conductors before and after lacquer coating; (b) Resistance measurement of lacquer-coated copper sheets using a micro-ohmmeter; (c) Conductivity test of copper powder; (d) Resistance of patterned copper sheets; (e) Performance comparison of copper sheets, copper tape, and copper foil.}
\label{fig:compatibilitytest}
\Description{This figure presents a comparative study of the electrical resistance of various conductive materials used in lacquer-integrated circuits.}
\end{figure*}

The unique decorative techniques of lacquer art are a key component of its artistic appeal. Building on common lacquer techniques (such as Miaoqi, Qiangjin, Jinyinpingtuo, luodian, and Xipi), we extend these traditional processes by seamlessly integrating embedded interactive circuits. Figure \ref{figure4} illustrates how Lacquer Art Interfaces can be realized using five common techniques. The process can be summarized in five main steps:

\begin{enumerate}
\item {\emph{Surface Preparation}}: Following traditional techniques, the lacquer surface is meticulously layered and polished multiple times to achieve a flawless, smooth finish.
\item {\emph{Circuit Design}}: Embedded circuits between layers, leaving extra length for connecting with the control unit.
\item {\emph{Surface Layer and Pattern Creation}}: Cover the circuits according to the specific lacquer technique used, and the desired patterns are created on the surface.
\item {\emph{Surface Finishing}}: Polishing ensures smoothness and full circuit embedding.
\item {\emph{Connection and System Adjustment}}: Connect the lacquerware interface to the control unit to enable interactive features and functionality.
\end{enumerate}

\subsection{Sensor performance}
To further illustrate how the design adapts to different craftsmanship needs, we summarize various sensor types, their ideal placement within lacquerware, and their compatibility with different lacquer techniques. Through a series of experiments, we tested the performance of capacitive, pressure, and temperature sensors within the lacquer interface, evaluating their functionality under varying lacquer layer thicknesses and their overall compatibility with the lacquer materials (Table \ref{tab:sensor_techniques}).

\subsubsection{Compatibility Test of Conductive Materials with Lacquer}

The goal of performance testing is to ensure effective capacitive sensors can be constructed without limiting the artist’s choice of materials. We evaluated the electrical conductivity of copper sheets, copper tape, brass foil, and copper powder across various widths and patterns. Based on common decorative line widths in traditional lacquer art, we created samples of each material at 1mm, 3mm, and 5mm widths, as well as 5mm wide patterned bands, and measured the resistance changes after embedding them in the lacquer layer (Figure \ref{fig:compatibilitytest}-a). Additionally, we assessed resistance under different conditions for various materials and widths (Figure \ref{fig:compatibilitytest}-e).

\emph{Results}: In term of material compatibility, all materials except copper powder showed no significant resistance changes after lacquer application, indicating that the lacquer does not affect their conductivity (Figure \ref{fig:compatibilitytest}-a, Figure \ref{fig:compatibilitytest}-b). Regarding material performance, copper powder exhibited unstable conductivity due to large particle gaps, making it unsuitable for interactive circuits (Figure \ref{fig:compatibilitytest}-c). Metal sheets with complex patterns maintained good conductivity (Figure \ref{fig:compatibilitytest}-d), enabling the integration of aesthetics and functionality. Additionally, As shown in Figure \ref{fig:compatibilitytest}-e, copper sheets and copper tape had resistance below 1$\Omega$ across all widths, while brass foil had slightly higher resistance, remaining below 2$\Omega$.

\begin{table}[H]
\centering
\setlength{\tabcolsep}{1.5pt}
\renewcommand{\arraystretch}{1.631} 
\caption{Correspondence between sensors and techniques}
\vspace{-2pt}
\footnotesize
\color{darkgray}
\begin{tabular}{lcccccc}
\toprule
\textbf{Sensor Type} & \textbf{Location} & \textbf{Miaoqi} & \textbf{Qiangjin} & \textbf{Jinyinpingtuo} & \textbf{Luodian} & \textbf{Xipi} \\ 
\midrule
Capacitive sensor & Surface & \checkmark & \checkmark & \checkmark &   & \checkmark \\ 
Pressure sensor & Inside & \checkmark &  & \checkmark & \checkmark &  \\ 
Temperature sensor & Inside & \checkmark &  & \checkmark & \checkmark &  \\ 
\bottomrule
\end{tabular}
\label{tab:sensor_techniques} 
\end{table}

\begin{figure*}[ht]
\centering
\includegraphics[width=1\textwidth]{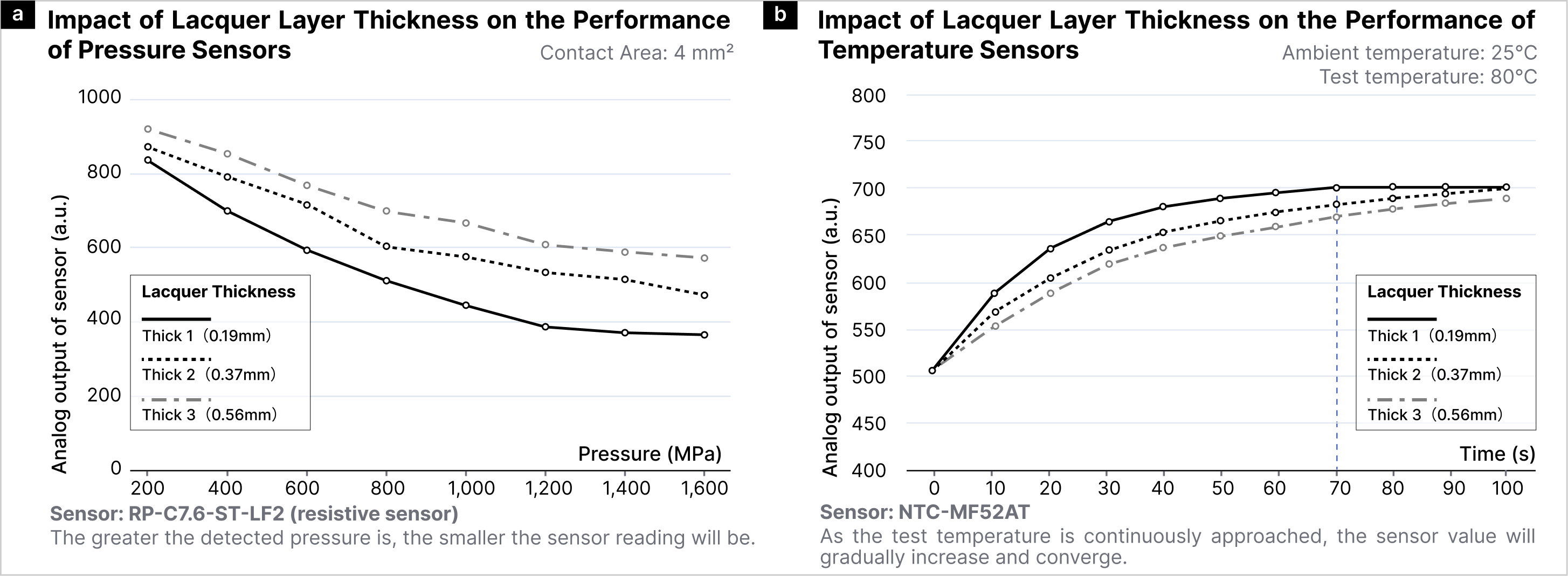}
\vspace{-0.5cm} 
\caption{Sensor performance with lacquer layers. (a) Pressure sensor; (b) Temperature sensor.}
\label{fig:sensor}
\Description{This figure illustrates the impact of varying lacquer layer thickness on the sensor output in pressure and temperature sensing scenarios.
(a) Impact on pressure sensors (RP-C7.6-ST-LF2, resistive type) – The graph shows a decrease in analog output as pressure increases, with different lacquer thicknesses (0.19 mm, 0.37 mm, and 0.56 mm) exhibiting varying response characteristics. Thicker lacquer layers result in higher resistance, reducing sensitivity to applied pressure.
(b) Impact on temperature sensors (NTC-MF52AT, thermistor type) – The graph depicts the sensor response as the test temperature increases from 25°C to 80°C. The analog output gradually rises and converges as the target temperature is approached. Different lacquer thicknesses influence the rate of response, with thicker coatings introducing a slight delay in sensor adaptation.}
\end{figure*}

\subsubsection{Concentration Test of Lacquer for Optimal Circuit Embedding}
In a controlled environment (28°C, 75\% RH), we tested three lacquer-to-oil (camphor oil, Fuzhou Dongchang Raw Paint Co., Ltd) ratios (1:0.8, 1:1, 1:1.2). These ratios were selected based on traditional lacquer practices, spanning from thinner to thicker applications. Each ratio was applied with thin, medium, and thick coatings (Figure \ref{fig:compatibilitytest}-e). Evaluations included: Levelness: Assessed surface flatness using a digital level. Layer Thickness: Measured each layer's thickness with a digital micrometer.

\begin{figure}[H]
\centering
\includegraphics[width=\linewidth]{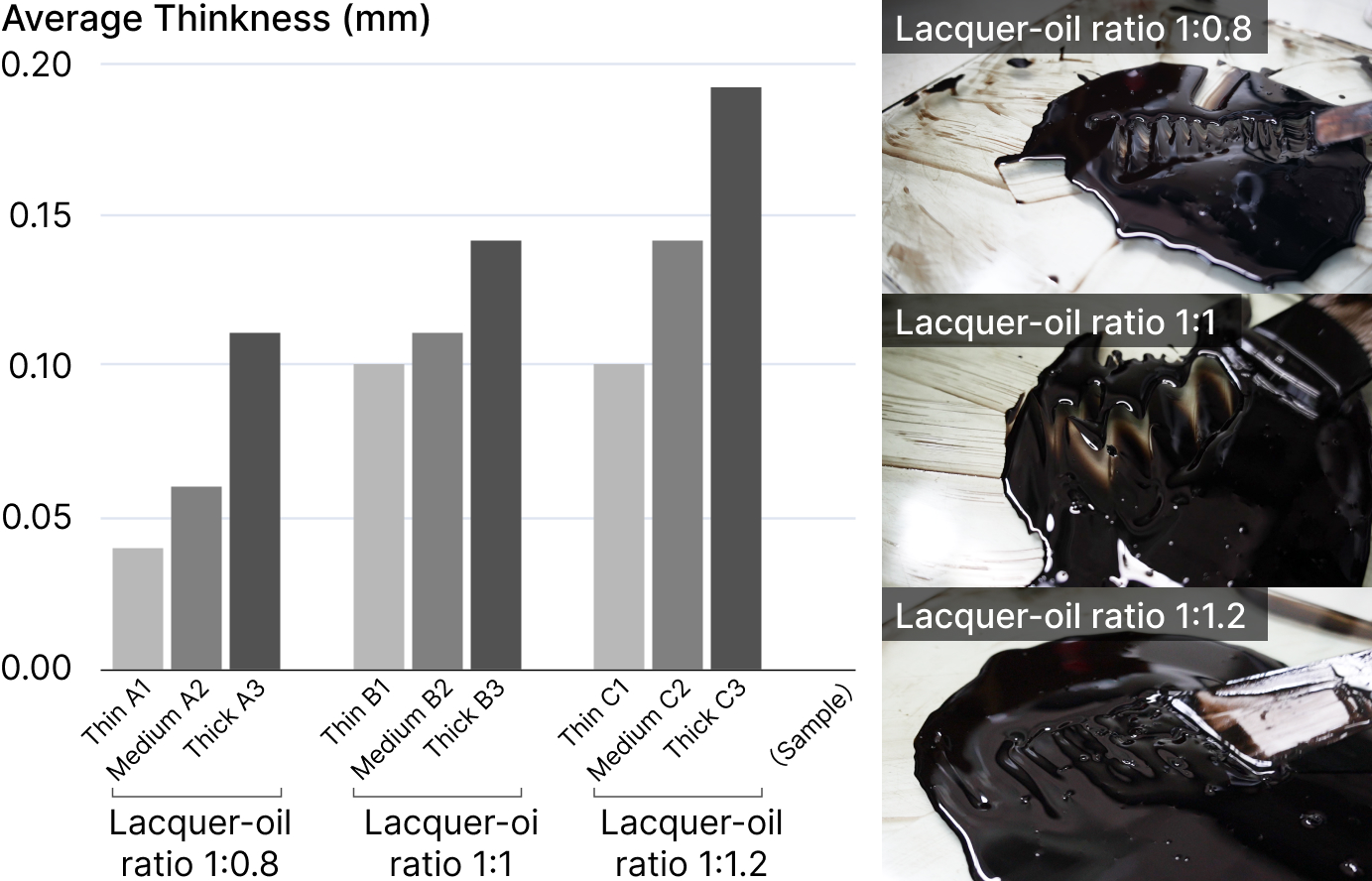}
\caption{Effect of lacquer-to-oil ratio on paint layer thickness.}
\label{fig:thinkness}
\vspace{-5mm}
\Description{This figure examines the effect of varying lacquer-oil ratios on the thickness of applied coatings.
Left: Average thickness measurements – The bar chart displays the measured lacquer layer thickness across different sample groups, categorized by lacquer-oil ratios (1:0.8, 1:1, and 1:1.2). Each ratio includes three different application thickness levels: thin, medium, and thick. As the lacquer-oil ratio increases, the overall thickness tends to rise, with the highest values observed at a 1:1.2 ratio (.
Right: Visual presentation of lacquer fluidity – Three images display the appearance and spread of lacquer mixtures under different lacquer-oil ratios, from top to bottom.}
\end{figure}

\emph{Results}: Figure \ref{fig:thinkness} illustrates the thickness changes under various lacquer-oil ratios and painting types. A lacquer-oil ratio of 1:1.2 applied with a thick brush produced the thickest and most consistent single layer (0.20mm) with optimal leveling (surface deviation <0.05mm). This combination effectively covers the circuit quickly,forming a uniform protective layer. This method provides a reliable foundation for manufacturing interactive lacquerware and was used in subsequent experiments.

\subsubsection{Performance Test of Sensors with Varying Lacquer Layer Thickness}
Since sensors need to be embedded within the lacquer layer, its thickness may impact their performance. This study evaluated the pressure sensor RP-C7.6-ST-LF2 (Figure \ref{fig:sensor}-a) and the temperature sensor NTC-MF52AT (Figure \ref{fig:sensor}-b) across lacquer thicknesses of 0.19mm, 0.37mm, and 0.56mm, assessing their responses under pressure and temperature conditions.

\emph{Results}: For the pressure sensor, when the same pressure was applied, detected pressure values decreased with increasing thickness (Thick3 < Thick2 < Thick1) (Figure \ref{fig:thinkness}). The temperature sensor in the thin lacquer layer (0.19mm) converged within 75 seconds, whereas thicker layers (0.37mm and 0.56mm) had longer convergence times, and some samples did not fully stabilize (Figure 7). These results indicate that thicker lacquer layers reduced sensor sensitivity. Therefore, it is essential to appropriately control lacquer thickness during the creation process to ensure the stability of interactive functionalities.

\begin{figure*}[htbp]
\centering
\includegraphics[width=\textwidth]{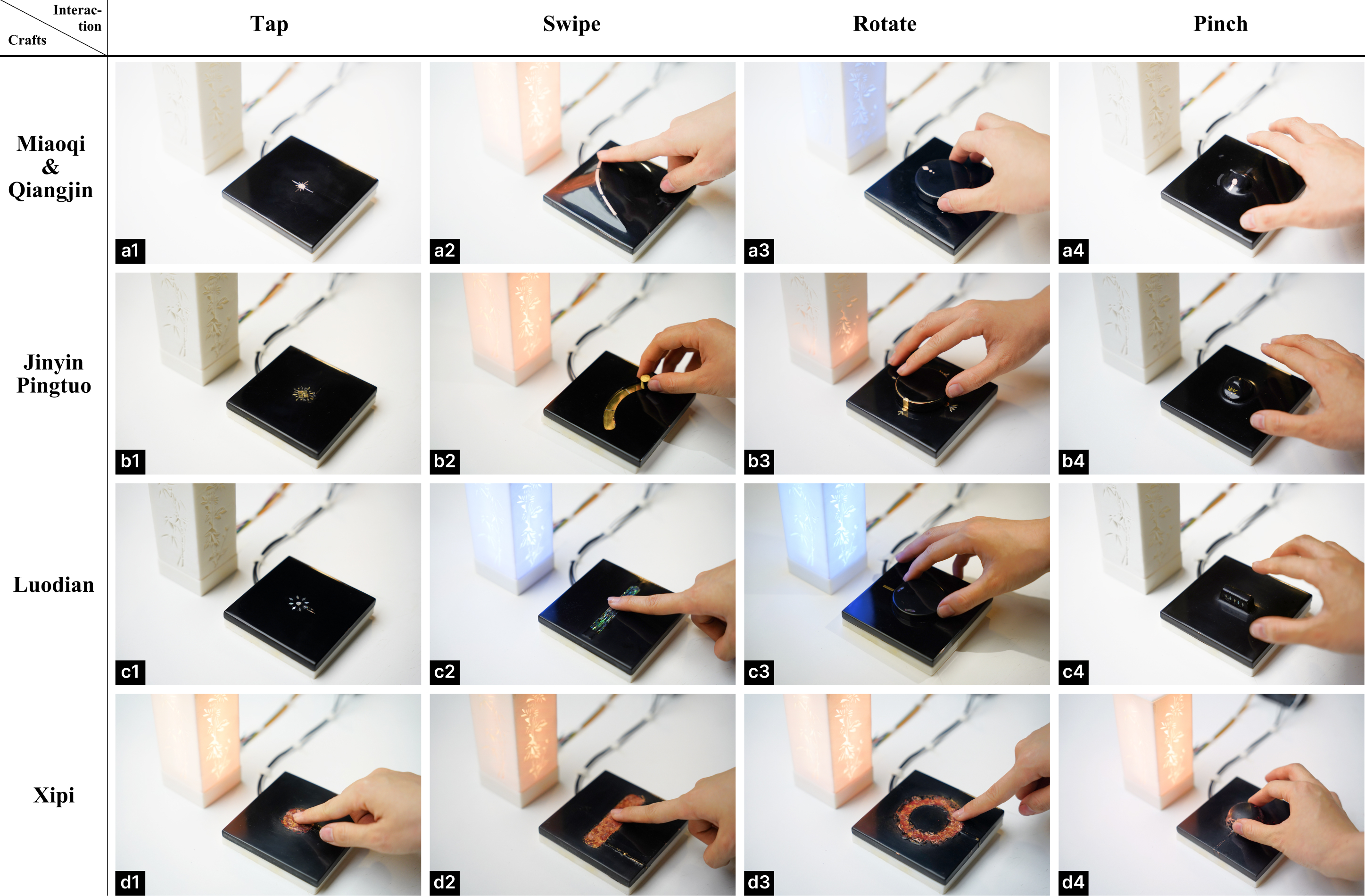}
\caption{Interaction primitives.}
\label{fig:primitives}
\Description{}
\end{figure*}

\subsection{Interaction Primitives.}

Although lacquer art is commonly used in daily life, its exploration in interactive design is still in its early stages, with its potential to bridge the physical and digital worlds yet to be fully realized. Through digital lacquer prototype experiments, we aim to explore the interactive possibilities of lacquer interfaces.

Building on the interaction framework by Foley and Shneider-man\cite{shneiderman1997direct}, we selected four intuitive user actions—tap, swipe, rotate, and pinch—aligned with common user habits. These actions were incorporated into a demo experiment with artisans using the five typical lacquer techniques (Figure \ref{fig:primitives}). These simple, everyday interactions proved reliable with circuits embedded in lacquer layers, demonstrating the feasibility of combining lacquer art with interactive technologies.

Lacquer's material compatibility and flexibility offer greater freedom in interface design, enabling a more natural user experience. For example, in demo B4, the interface was centered around the pinch gesture, allowing users to adjust digital parameters in real time through hand movements. Traditional lacquer patterns further enhance interaction, as seen in demo C2, where raden (inlaid shell) patterns reflected changes in brightness, aligning physical and digital perception for a richer sensory experience.

Lacquer interfaces support customizable interactions while preserving traditional aesthetics. Even simple functions like lighting control can be personalized through different techniques and patterns, showcasing the seamless integration of lacquer art and digital technologies and opening new possibilities for incorporating HCI into everyday life.

\section{Toolkit Design}
To promote the adoption of Layered Interactions and foster bidirectional learning and collaboration between craftsmen and technology specialists, we developed a toolkit and conducted an experimental study involving both groups. The study aimed to explore participants' perspectives on craft-centric design strategies, the toolkit provided, and the collaborative workflow outlined in the "Craft-Tech Alignment Manual". Section 4 outlines the initial toolkit design and refinement process, including a pilot experiment with two craftsmen and two HCI researchers to assess usability and guide further optimization. Building on this, Section 5 presents insights from seven craftsmen and nine HCI researchers using the toolkit in creative endeavors. Table \ref{tab:2satge} illustrates the procedural flow of both experimental phases.

\begin{table*}[ht]
\color{darkgray}
\centering
\caption{Interactive development and evaluation of toolkit}
\vspace{-3mm}
\fontsize{8.7}{11.4}\selectfont
\setlength{\tabcolsep}{2.2mm}{\begin{tabular}{lllll}
\textbf{Stage} & \textbf{Description}  & \textbf{Participants} & \textbf{Output} & \textbf{Goal/Significance}                                \\ \hline

\renewcommand{\arraystretch}{1.2}{\multirow{2}{*}{\begin{tabular}[c]{@{}l@{}}
\\Stage 1\end{tabular}}}
        & \begin{tabular}[c]{@{}l@{}}Created the initial\\ version of the toolkit\end{tabular} & Authors and craftsman                        & \begin{tabular}[c]{@{}l@{}}Preliminary Toolkit (version 1.0) \end{tabular}                                                       & \begin{tabular}[c]{@{}l@{}}Establish a baseline toolkit\\for initial user testing\end{tabular}              \\ \cline{2-5} 
        & \begin{tabular}[c]{@{}l@{}}Experiment 1: Optimize\\ the toolkit\end{tabular}                         
        & \begin{tabular}[c]{@{}l@{}}{[}A1/B1{]} Dynamic lacquer painting\\ {[}A2/B2{]} Contextualized chess\end{tabular} 
        & \begin{tabular}[c]{@{}l@{}} layered Interactions \\Design Toolkit (version 2.0)\end{tabular}                                          & \begin{tabular}[c]{@{}l@{}}Identify usability gaps and\\improve functionality and\\versatility\end{tabular}  
\\ \hline
Stage 2                                      
        & \begin{tabular}[c]{@{}l@{}}Experiment 2: Use\\ version 2.0 to create\\ practical works\end{tabular}  & 16 participants in 5 groups                 
        & \begin{tabular}[c]{@{}l@{}}6 lacquer interactive artworks\end{tabular} 
        & \begin{tabular}[c]{@{}l@{}}Validate version 2.0's\\ adaptability and usability\\ in artistic tasks\end{tabular} 
\\ \hline
\end{tabular}}
\label{tab:2satge}
\vspace{-1mm}
\end{table*}

\subsection{Preliminary Toolkit}
Based on an analysis of lacquer craft's characteristics and processes and the collaboration with lacquer craftsmen in design primitives phase, an initial version of the toolkit was developed. A preliminary design outline was constructed that aligns with the structure, procedures, and technological possibilities inherent in lacquer artistry, mapping traditional craft steps with interactive technologies individually.

To quickly evaluate the toolkit's usability, a small-scale experiment was conducted with two groups: Group A (A1 \& B1) and Group B (A2 \& B2). Before the experiment, we briefed the participants on the functionalities and operational methods of the toolkit modules and supplied the requisite materials for their creations. Each group, comprising a lacquer artisan (A1, A2) and an HCI researcher (B1, B2), collaborated to complete the entire process of designing and fabricating interactive lacquer art within 48 hours. After the experiment, each group conducted a 20 minutes semi-structured interviews to collect feedback.

\begin{figure}[H]
\centering
\includegraphics[width=\linewidth]{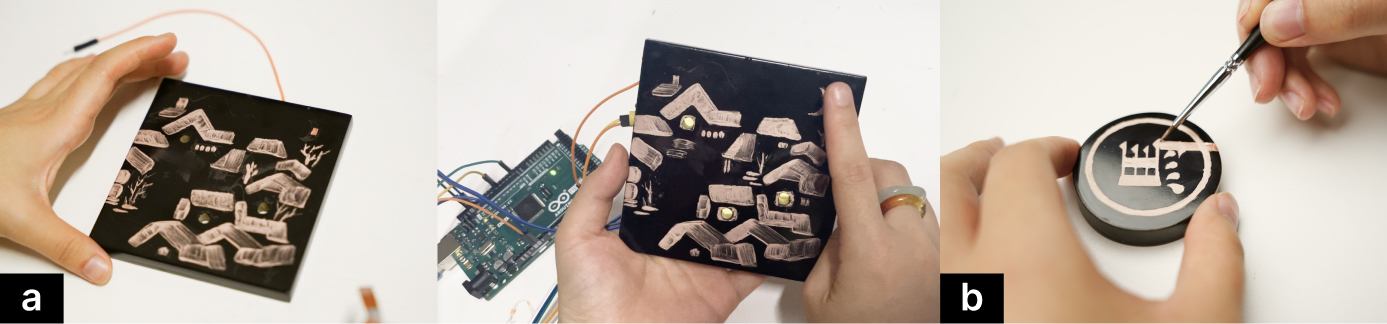}
\vspace{-4mm}
\caption{Digital lacquer artworks using the initial toolkit: (a) Dynamic Lacquer Painting - Moonlight: Touching the moon lights up countless home; (b) Contextualized Chess: Pieces trigger sounds, enhancing realism and immersion.}
\label{fig:initial}
\Description{This figure shows the outcomes of a small-scale experiment evaluating the initial version of the interactive lacquer toolkit.
(a) Dynamic Lacquer Painting – "Moonlight" – Participant A1 painted a moonlit scene using Miaoqi techniques. The figure shows that when the moon motif is touched, embedded LEDs illuminate, symbolizing the sentiment of "lighting up countless homes."
(b) Contextualized Chess – Designed by participants A2 and B2, this interactive chess set enhances gameplay through auditory feedback.}
\end{figure}

A1 \& B1 crafted a dynamic interactive lacquer painting "Moonlight" (Figure \ref{fig:initial}-a). when the moon motif was touched, LED lights glow, evoking the sentiment of "illuminating thousands of homes." A2 \& B2 developed a sonically interactive chess set named "Sound Chess," (Figure \ref{fig:initial}-b). When the pieces make contact, they produce combative sounds, such as the "pawns" generating a clashing sound and the "knights" emitting a neighing sound. This feature enhances the immersive experience of the game. The findings revealed that the operational instructions to be relatively clear(A1 \& A2), yet there was an absence of explicit guidance on the modifiability of certain modules, which impacted the creative flexibility. Notably, A1 encountered confusion regarding the input-output mapping when fine-tuning the artwork's effects. B1 and B2 requested more comprehensive manuals that elucidate the interrelations among modules. Additionally, B2 noted that the multitude of interfaces heightened cognitive load and recommended a more streamlined design. In response to this feedback, we refined the toolkit, leading to the release of version 2.0, with enhancements such as:

\leftskip 10pt

\emph{Manual Enhancement}: Introduction of a new section "Correspondence between Technology and Craft" that delineates which modules are modifiable and which are not, along with clarifying input-output mapping relationships (e.g., code modules, input-output correlations, and interface details).

\leftskip 0pt

\begin{figure}[H]
\centering
\vspace{-3mm}
\includegraphics[width=\linewidth]{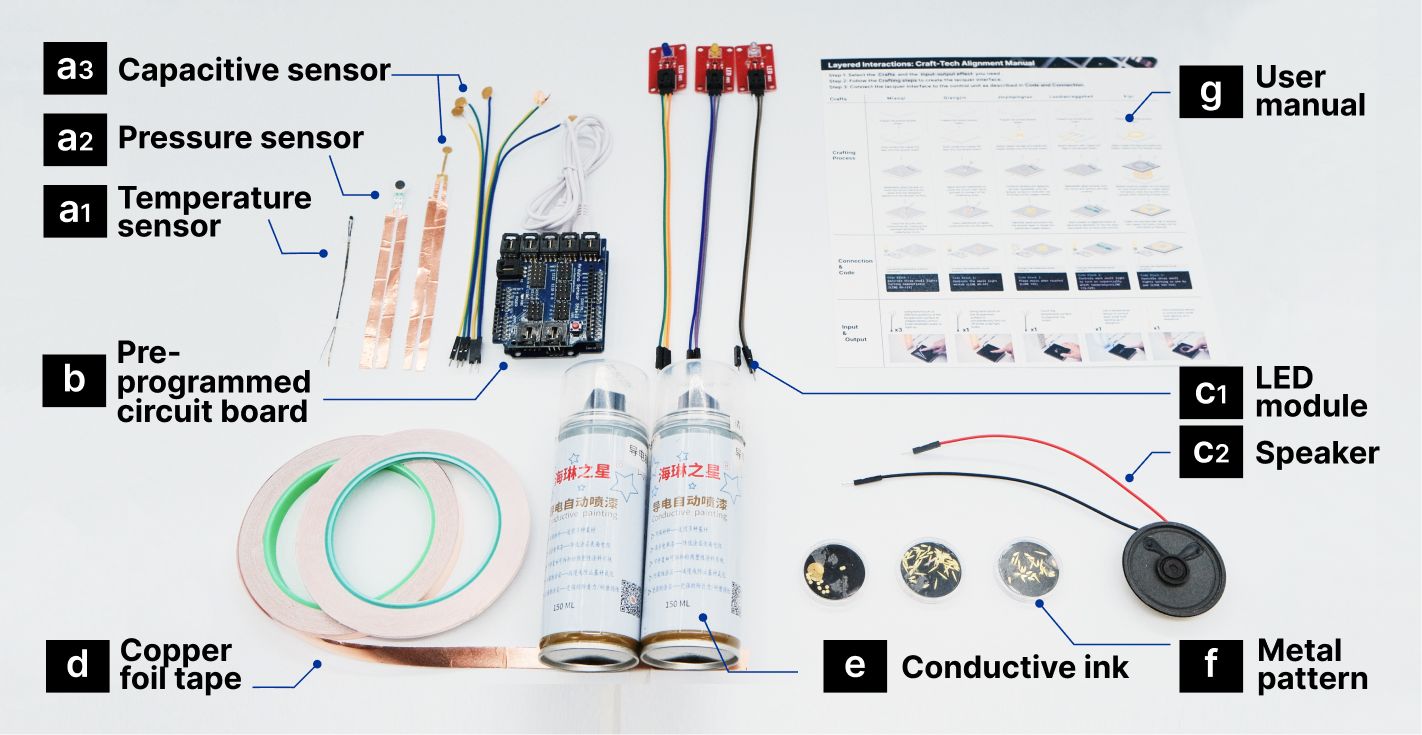}
\vspace{-4mm}
\caption{Layered Interaction Design Toolkit Components.}
\label{fig:toolkit}
\Description{Layered Interaction Design Toolkit Components.}
\end{figure}

\leftskip 10pt

\emph{Streamlining Interface Connection Guidance}: The reduction of superfluous interfaces to alleviate cognitive burden and enhance modularity.

\emph{Code Refinement}: The provision of additional annotations and exemplar code snippets to further diminish the technical learning threshold for HCI researchers.

\leftskip 0pt

\subsection{Layered Interactions Design Toolkit}
\subsubsection{Design Principles}
Existing toolkits often require basic training in circuitry principles \cite{10.1145/3532106.3533535}. During our interactions with traditional craftsmen, we found that electronic components and programming code often feel intimidating, which may cause them to abandon the tools. Thus, we focused on aligning the toolkit with the cognitive patterns and workflows of craftsmen, while simplifying the manipulation of electronic circuits. The toolkit is designed with the following key features:
\begin{enumerate}
    \item {\emph{Simplicity}}: It features a low learning curve, making it accessible to individuals with limited exposure to programming or electronics.
    \item {\emph{Intuitiveness}}:   It is designed to align with craftsmen's established design habits, facilitating a smoother integration of electronic technologies.
    \item {\emph{Customizability}}: It fosters the freedom for individualized creation, ensuring that the craftsman's creative expression and adaptability are not compromised.
\end{enumerate}

\subsubsection{Components of the Layered Interactions Design Toolkit}
The toolkit comprises the following components (Figure \ref{fig:toolkit}):
\begin{enumerate}
    \item {\emph{Input Components}}: Temperature sensor (a1), pressure sensor (a2), and capacitive sensor (a3).
    \item {\emph{Output Components}}: LED module (c1), speaker (c2), and a circuit board pre-loaded with code (b), as well as extension modules for the capacitive sensor (d, e, f). The integrated Arduino circuit board contains five distinct code modules. By following the manual's instructions, craftsmen can connect the sensors to the appropriate serial ports to achieve a range of interactive functionalities. The code is extensively annotated, delineating the sections open to modification and offering exemplars of such alterations, allowing customization for creative requirements.
    \item {\emph{User Manual (g)}}: This manual provides detailed instructions for integrating of interactive circuits within artworks connecting sensors, and activating their respective functions in line with the crafting process.
\end{enumerate}

\begin{figure*}[t] 
\centering
\includegraphics[width=\textwidth]{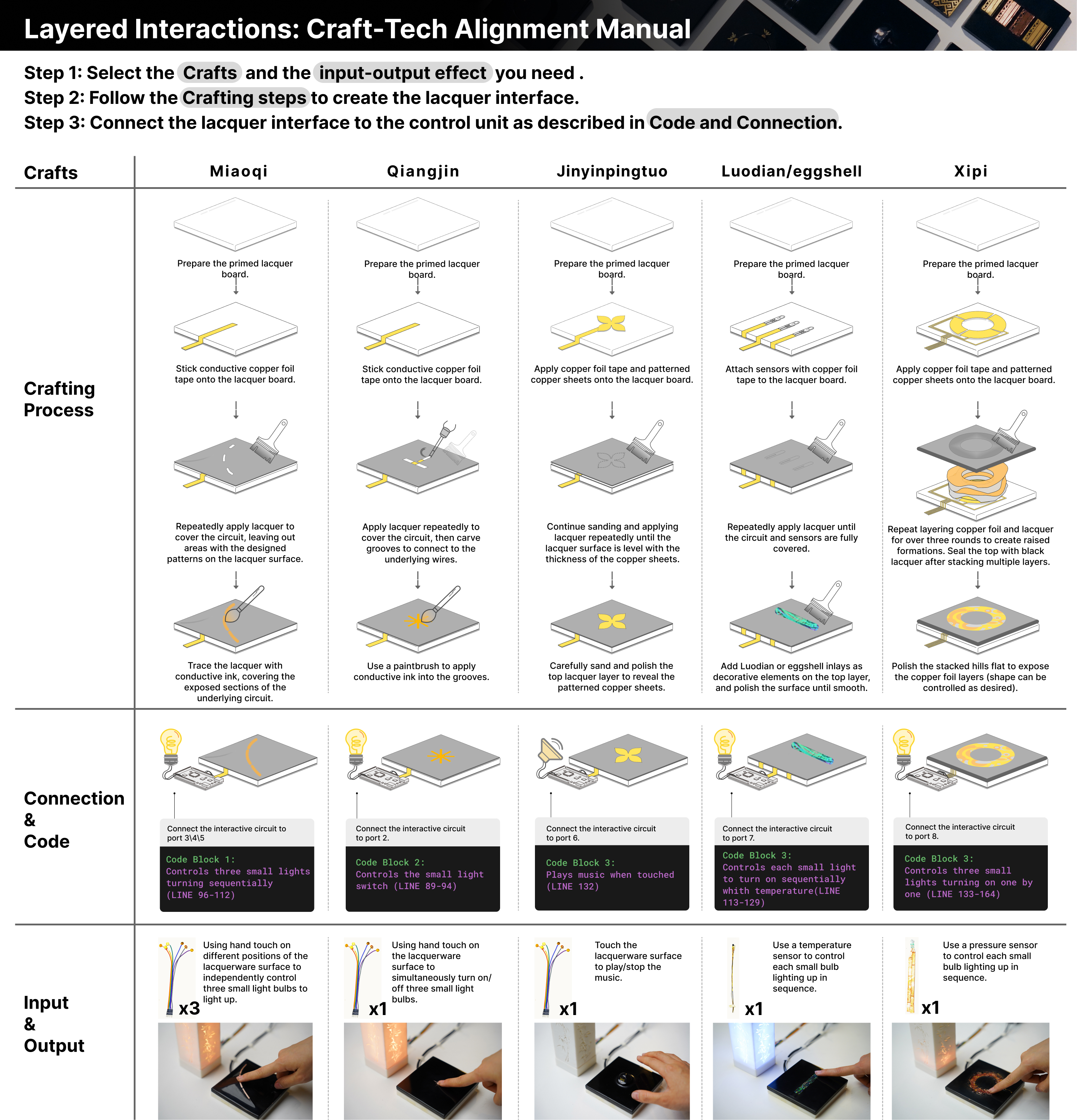}
\caption{Layered interactions: Craft-tech alignment manual.}
\label{fig:manual}
\Description{Layered interactions: Craft-tech alignment manual.}
\end{figure*}

\subsubsection{Craft-Tech Alignment Manual}
Demonstration phase observations show that artists prioritize the final output effects over the principles of the circuitry. Consequently, we have developed the Craft-Tech Alignment Manual to serve as a directive instrument (Figure \ref{fig:manual}). The manual is structured with the upper section cataloging the lacquer techniques supported and their interfacing procedures with interactive circuits; the central portion elucidates the methods of wiring and the selection of corresponding code; the lower section visually presents the achievable effects and the requisite electronic components.

\begin{table}[H]
\color{darkgray}
    \centering
    \setlength{\tabcolsep}{4.3pt} 
    \caption{Background of participants}
    \vspace{-2mm}
    \footnotesize
    \renewcommand{\arraystretch}{1.5} 
    \arrayrulecolor{darkgray} 
    \begin{tabular}{lll}
    \toprule
     \textbf{Ppts} & \textbf{Degree} & \textbf{Description} \\
    \midrule[0.3pt]
     A1 &  Graduate Student &  Lacquer Art, Glass Art \\
     A2 &  Graduate Student &  Digital Craft (2 yrs exp. in Lacquer Art)\\
     A3 &  PhD Candidate &  Lacquer Art \\
     A4 &  PhD Candidate &  Lacquer Art\\
     A5 &  Graduate Student &  Lacquer Art \\
     A6 &  PhD Candidate &  Sculpture (1 yr exp. in Lacquer Art)\\
     A7 &  Undergraduate Student &  Lacquer Art \\
     B1 &  Master's Student &  HCI (Science Popularization Products) \\
     B2 &  Master's Student &  HCI (Product
     Interaction Design) \\
     B3 &  PhD Candidate &  HCI (Interaction Design for Children) \\
     B4 &  Master's Student &  HCI (Science Popularization Products) \\
     B5 &  PhD Candidate &  HCI (AI Interaction Design) \\
     B6 &  Graduate Student &  HCI (Integrated Circuits) \\
     B7 &  Master's Student &  HCI (Electronics and Information Engineering) \\
     B8 &  Research Scientist &  HCI (Recommender Systems) \\
     B9 &  PhD Candidate &  HCI (Computer Science) \\
    \bottomrule[0.3pt]
    \end{tabular}
    \label{tab:participants}
    \Description{}
\end{table}

\section{Collaborative Attempt and Verification}
\subsection{Methodology}
To explore more effective methodologies for supporting the creative practices of lacquer interfaces, we convened a collaborative creation session with 7 artists (all with a foundation in lacquer art) and 9 digital technology experts (Table \ref{tab:participants}). Artists are labeled as Group A, while the digital technology specialists are labeled as Group B. The experiment involved 16 participants, allocated into six groups. Each group included at least one artist and one technical expert, and the formation of groups considered the expertise level in their respective domains. Four authors were assigned to the groups to observe and provide minimal assistance if the work reached an impasse. Artists employed the \textit{layered Interactions Design Toolkit} for their creations, with HCI researchers contributing as collaborators in ideation and providing assistance with circuit assembly, culminating in the co-creation of comprehensive interactive lacquer artworks. The experimental protocol included four principal components:

\begin{enumerate}
    \item {\emph{Introduce Layered Interactions}}: Providing an overview of lacquer material properties, integration mechanisms and underlying principles, toolkit design, supported artisanal practices, and output effects.
    \item {\emph{Interview on Creative Habits}}: Artists detailed their personal crafting routines for both simple lacquer paintings and intricately shaped works.
    \item {\emph{Toolkit Usability}}: Participants collaborate in ideation, circuit design, lacquer application and pattern creation, code alteration, and final touches.
    \item {\emph{Assessment and Feedback}}: (a) Likert Scale Application: The scale (1 = strongly disagree, 5 = strongly agree) measured the toolkit's user-friendliness, utility, and the efficacy of cross-disciplinary collaboration. (b) 50 minutes Semi-Structured Interviews: Conversations revolved around toolkit usability evaluation, the Craft-Tech Alignment Manual, exploring its role in facilitating collaboration and innovation, and discussing the craft-centered non-intrusive approaches. The entire interview was recorded and transcribed verbatim. Two researchers performed initial open coding on the data, which was followed by a collaborative discussion to resolve any coding divergences, ensuring the analysis's reliability and compatibility.
\end{enumerate}

\begin{table*}[ht]
\centering
\setlength{\tabcolsep}{8pt}
\caption{Details of collaborative products}
\vspace{-1.4mm}
\small
\color{darkgray}
\begin{tabular}{ccll}
\toprule
\textbf{Artifact} & \textbf{Figure} & \textbf{Toolkit Usage} &  \textbf{Participants} \\

\midrule
 a.Musical tableware             & \begin{minipage}[b]{0.46\columnwidth}\raisebox{-.4\height}{\includegraphics[width=\linewidth]{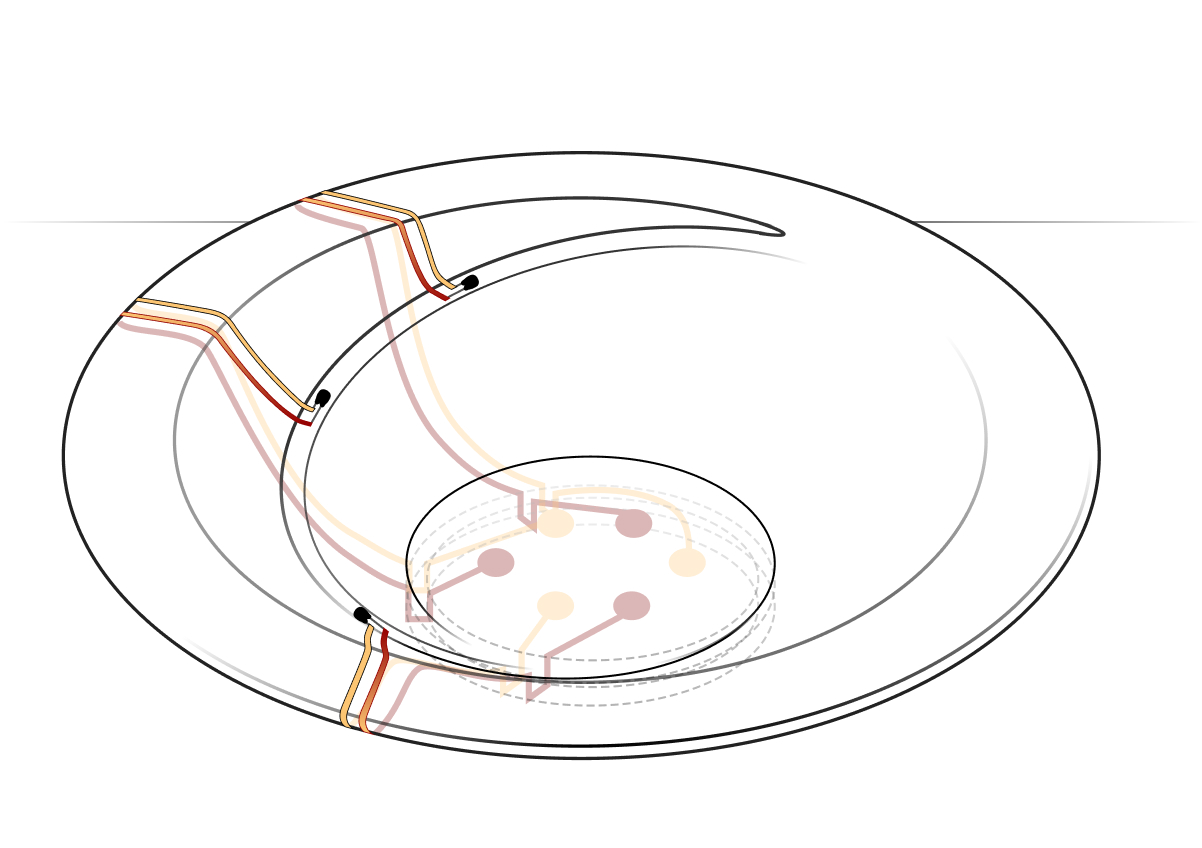}}\end{minipage} &  \begin{tabular}[c]{@{}l@{}}\textbf{Input}: 3 temperature sensors arranged along grooves\\and covered with lacquer\\ \textbf{Output}: detecting temperature changes triggers\\musical performance\\ \textbf{Code}: code block 3\&4 (with a few adjustments)\end{tabular} &  \begin{tabular}[c]{@{}l@{}}A1, A7, B1, B2\\Author 1\end{tabular} \\

 b.Glow-transition lamp              & \begin{minipage}[b]{0.46\columnwidth}\raisebox{-.4\height}{\includegraphics[width=\linewidth]{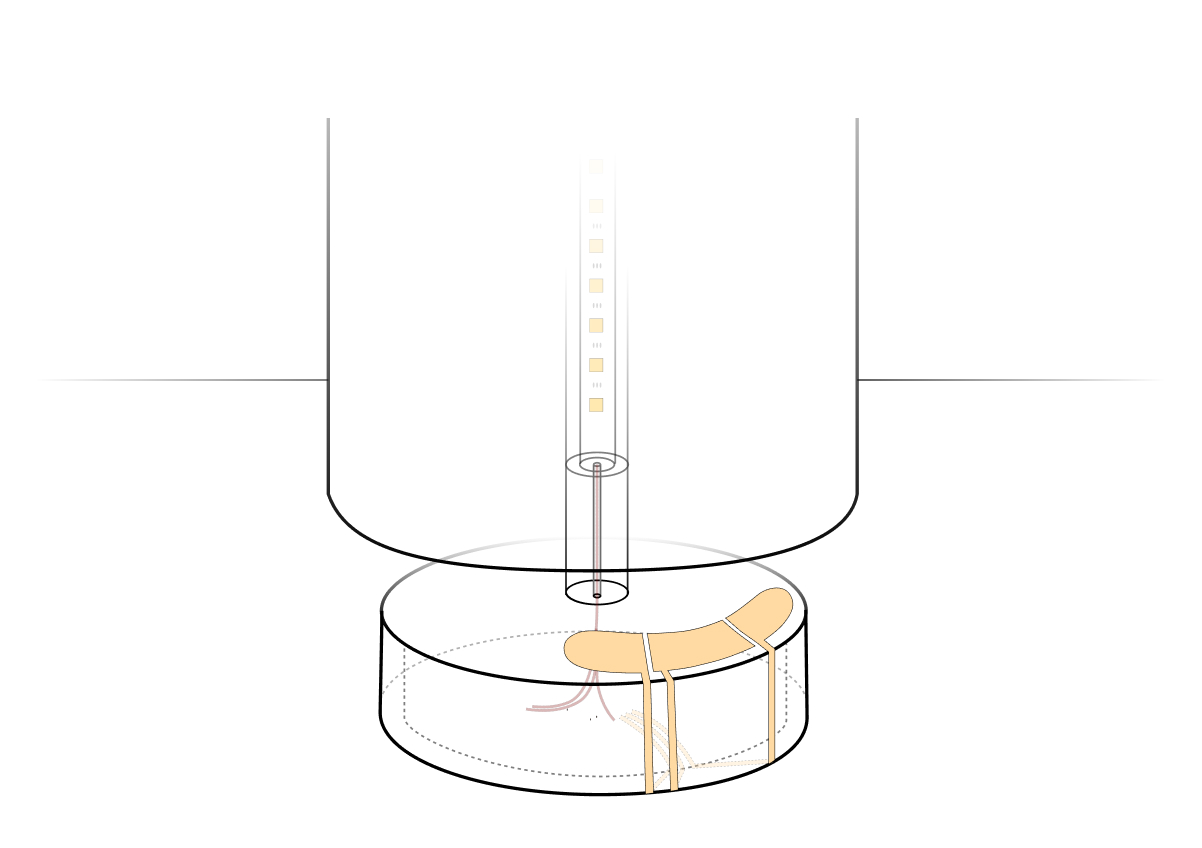}}\end{minipage} &  \begin{tabular}[c]{@{}l@{}}\textbf{Input}: 3 capacitive sensors made from copper sheets\\ \textbf{Output}: sliding from left to right gradually\\increases brightness\\ \textbf{Code}: code block 1\end{tabular} &  \begin{tabular}[c]{@{}l@{}}A2, B7\\Author 2 \end{tabular} \\

 c.Interactive cat figurine               & \begin{minipage}[b]{0.46\columnwidth}\raisebox{-.4\height}{\includegraphics[width=\linewidth]{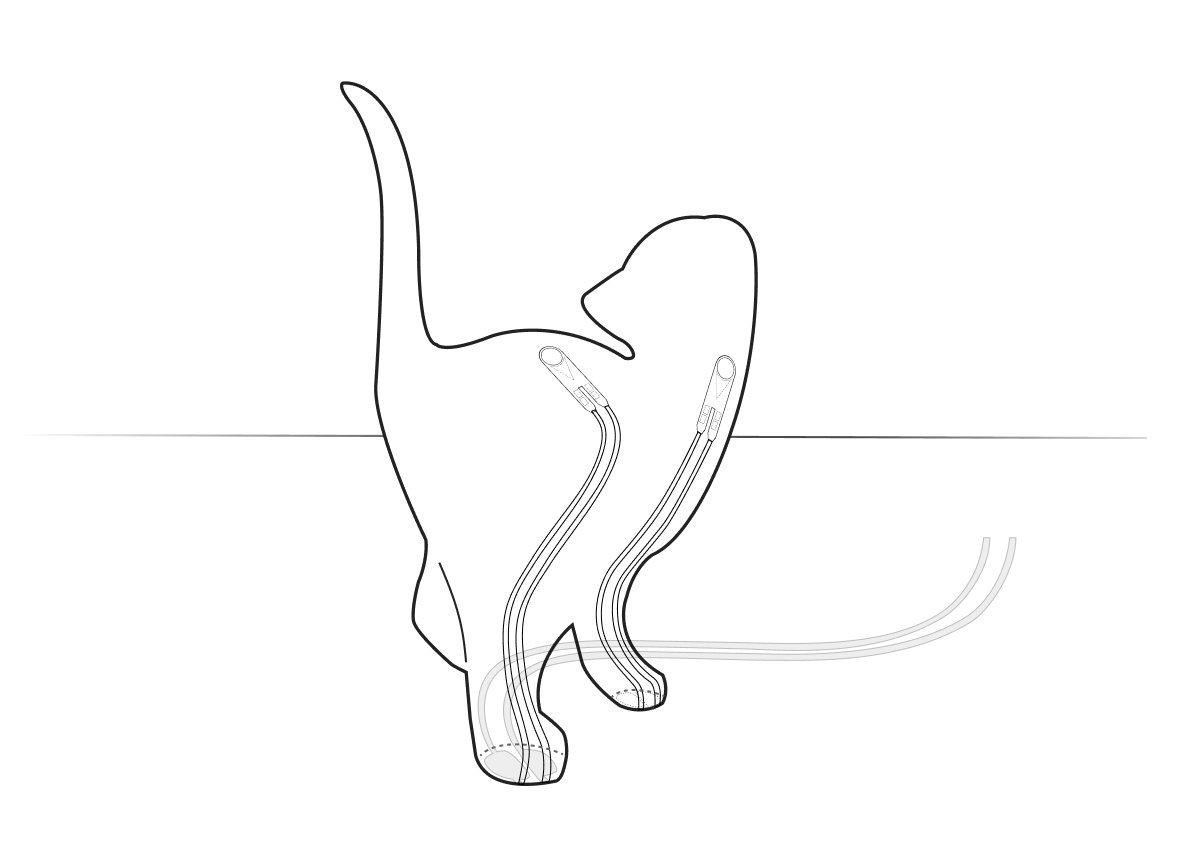}}\end{minipage} &  \begin{tabular}[c]{@{}l@{}}\textbf{Input}: 2 pressure sensors placed on the neck and back\\ \textbf{Output}: stroking these areas produres different cat\\sounds, and the screen highlights the touched region\\ \textbf{Code}: code block 3\&5 (with a few adjustments)\end{tabular} &  \begin{tabular}[c]{@{}l@{}}A3, B3, B8\\Author 3\end{tabular} \\

 d.Rhythm-responsive bracelet          & \begin{minipage}[b]{0.46\columnwidth}\raisebox{-.4\height}{\includegraphics[width=\linewidth]{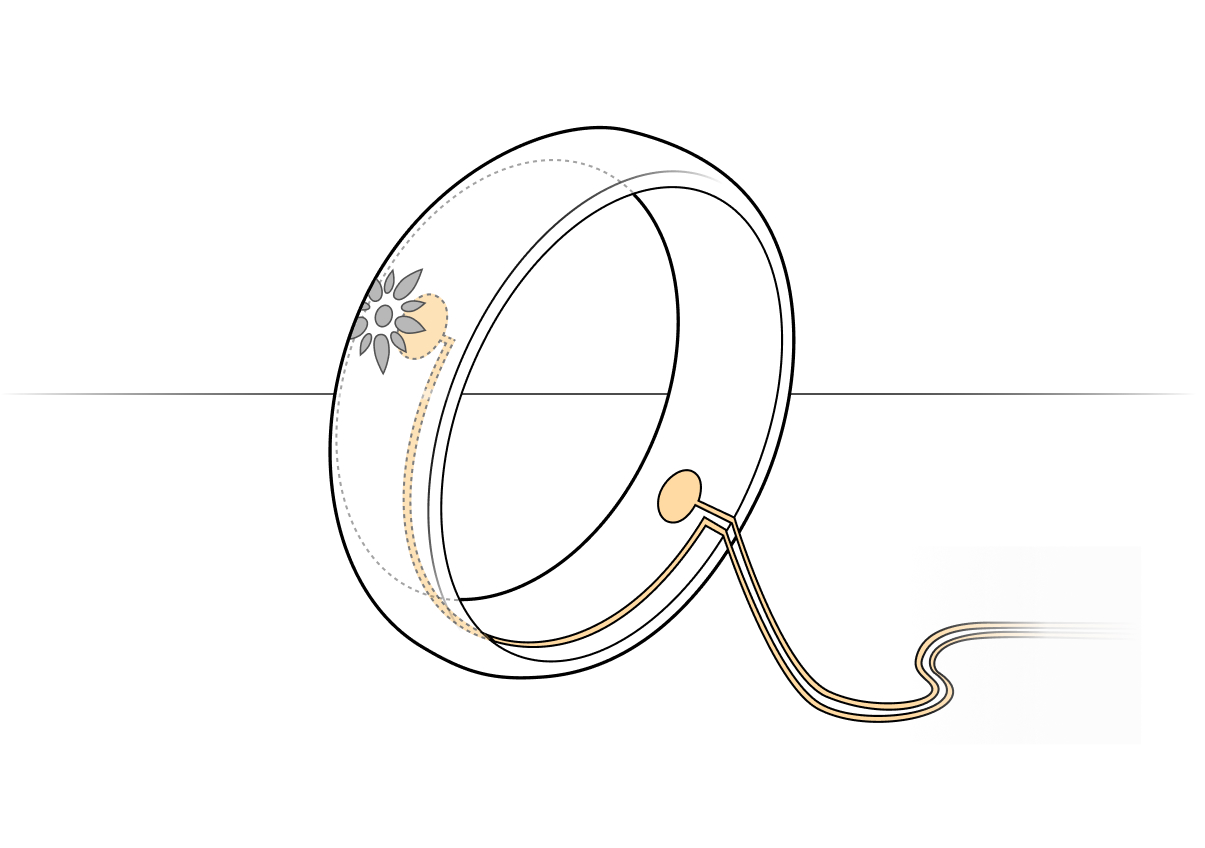}}\end{minipage} &  \begin{tabular}[c]{@{}l@{}}\textbf{Input}: 2 capacitive sensors made from copper sheets\\ \textbf{Output}: touching the copper sheets  produces sound\\ \textbf{Code}: code block 3\end{tabular} &  \begin{tabular}[c]{@{}l@{}}A4, B4, B9\\Author 4\end{tabular} \\

 e.Smart lacquer ring              & \begin{minipage}[b]{0.46\columnwidth}\raisebox{-.4\height}{\includegraphics[width=\linewidth]{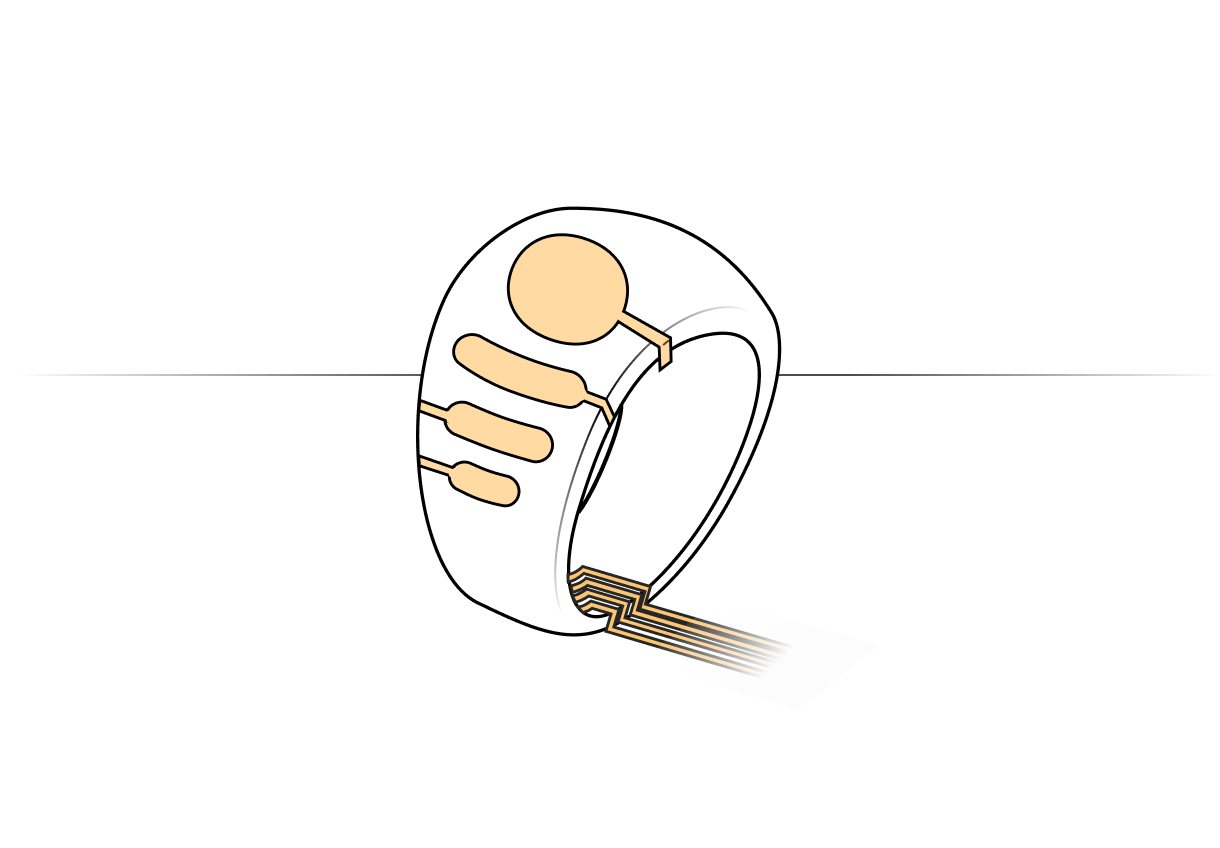}}\end{minipage} &  \begin{tabular}[c]{@{}l@{}}\textbf{Input}: 4 capacitive sensors made from copper sheets\\ \textbf{Output}: 3 ovals for control volume, while a circular\\copper  sheet toggles play/pause\\\textbf{Code}: code block 3 (with a few adjustments)\end{tabular} &  \begin{tabular}[c]{@{}l@{}}A5, B5\\Author 1 \end{tabular} \\

 f.Braille tactile interface & \begin{minipage}[b]{0.46\columnwidth}\raisebox{-.4\height}{\includegraphics[width=\linewidth]{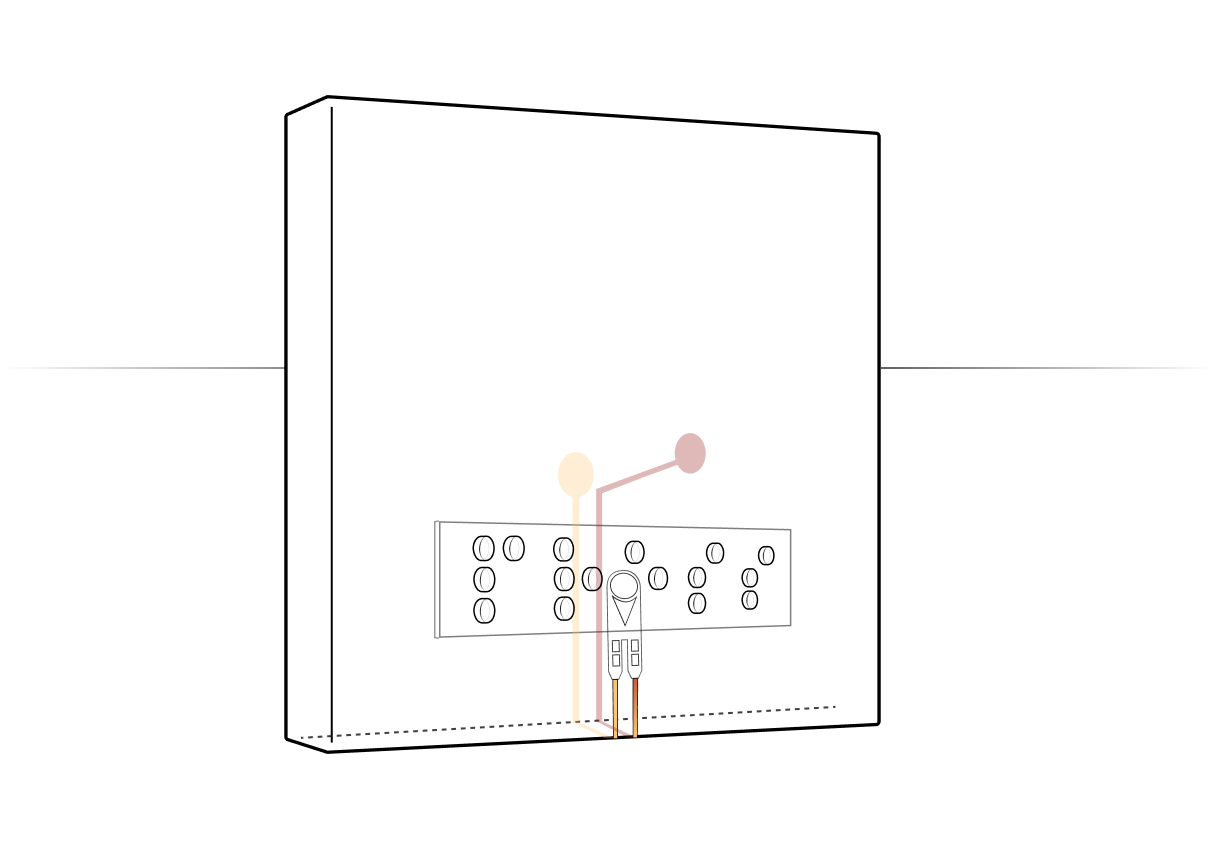}}\end{minipage} &  \begin{tabular}[c]{@{}l@{}}\textbf{Input}: 1 pressure sensor located beneath Braille\\ \textbf{Output}: Press the Braille; the greater the pressure,\\the brighter the light. \\\textbf{Code}: code block 5 (with a few adjustments)\end{tabular} &  \begin{tabular}[c]{@{}l@{}}A6, B6\\Author 2\end{tabular} \\
\bottomrule
\end{tabular}
\label{group}
\Description{}
\vspace{2mm}
\end{table*}

\subsection{Artworks of Collaborative Creation}
Six groups used the \textit{Layered Interactions Design Toolkit} to design six digital lacquer art products. These works illustrate how to imbue traditional lacquer with new functionalities without compromising its aesthetic and utility, demonstrating its versatile application potential in contemporary settings. Figure \ref{fig:cowork process} highlights one of the co-creations of a lacquerware artwork, demonstrating the collaborative steps and integration of digital technologies with traditional lacquer craftsmanship.

\subsubsection{Musical Tableware}
\begin{figure*}[htbp]
  \includegraphics[width=\textwidth]{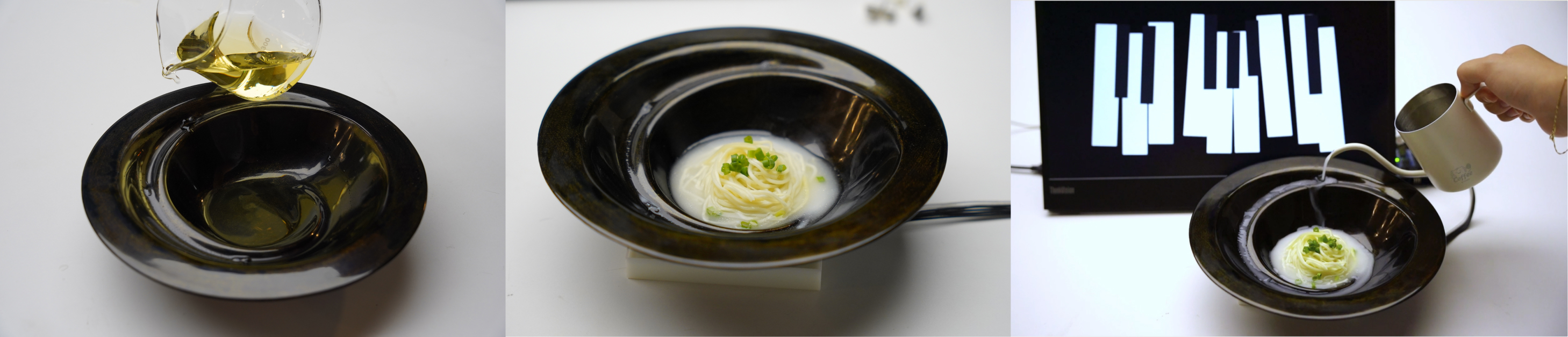}
  \vspace{-4mm}
  \caption{Digital lacquerware product: Musical tableware.}
  \vspace{-2mm}
  \label{fig:cowork1}
  \Description{Digital lacquerware product: Musical tableware.}
\end{figure*}

A1\&A7 noted that tableware are a quintessential application for lacquer, where patterned tableware and food combine to create the art of gastronomy. Group 1 designed a spiral-structured lacquer food vessel (Figure \ref{fig:cowork1}). A1\&A7 initially conceptualized the bowl's shape, B1 assisted modeling and 3D printing, A1 integrated a temperature sensor within the curved groove (Table \ref{group}-a), positioning sensors at the upper, middle, and lower flow paths of the water. The circuitry was discreetly placed at the bowl's base, reinforced with copper sheets for ease of movement and cleaning.

Group 1 applied the lacquer interface to the presentation process of gourmet food. As soup flows along the spiral grooves, hidden sensors detect temperature changes and trigger a musical response, transforming the dining experience into a multi-sensory interactive performance. This design retains the functionality of traditional tableware while introducing dynamic interactive elements, expanding lacquer’s role in modern life.

\subsubsection{Glow-transition Lamp}
A2\&B7 Utilizing lacquer's exceptional adhesion properties to experiment with the digital transformation of daily objects. They coated the base of lamp with lacquer and designed an interactive area using the interactive circuit with Xipi craft. The gradual color transition from dark to red visually symbolizes the interaction from darkness to light (Figure \ref{fig:cowork2}-a), delivering a direct and naturally pleasing interactive experience. A2 replaced the toolkit's small copper sheet and soldered the larger copper sheet onto the conductive tape to support the interactive area in their design (Table \ref{group}-b). B7 adjusted the appropriate lighting brightness, and collaborated with the artisan during the polishing stage of the Xipi technique to expose the necessary interactive touch point. This design demonstrates the potential of lacquer interfaces for seamless integration with existing products in future smart home environments, expanding the aesthetic boundaries of daily objects and enhancing diverse user experiences in home settings.

\subsubsection{Interactive Cat Figurine}
\begin{figure*}[htbp]
  \includegraphics[width=\textwidth]{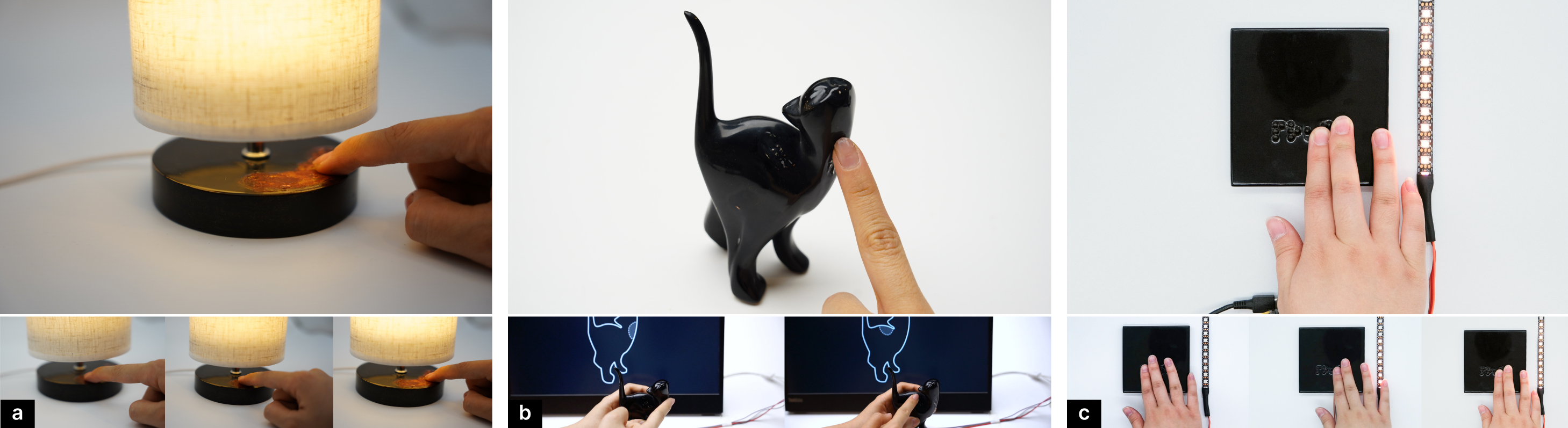}
  \vspace{-4mm}
  \caption{Digital lacquerware product. (a) Glow-transition lamp; (b) Interactive cat figurine; (C) Braille tactile interface.}
  \vspace{-2mm}
  \label{fig:cowork2}
  \Description{Digital lacquerware product. (a) Glow-transition lamp; (b) Interactive cat figurine; (C) Braille tactile interface.}
\end{figure*}

A3, B3\&B8 designed an emotionally resonant cat ornament (Figure \ref{fig:cowork2}-b). Two pressure sensors were installed on the cat's neck and back, activating the sensors hidden beneath the lacquer surface upon stroking these areas and triggering sound feedback (Table \ref{group}-c). For example, gently stroking the cat’s neck elicits a soft purring sound, while stroking its back may trigger a happy meow, allowing users to experience an emotional connection similar to interacting with a real cat. The addition of interactive elements transforms the figurine from a static decorative object into a dynamic medium for emotional expression.

\begin{figure*}[htbp]
\centering
\includegraphics[width=\textwidth]{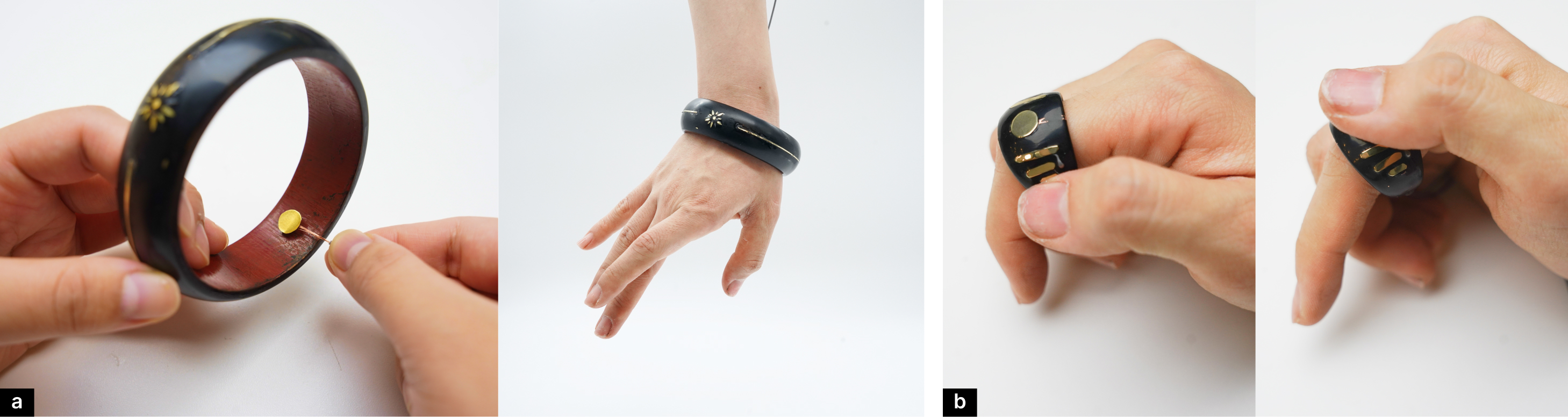}
\vspace{-4mm}
\caption{Digital lacquerware product. (a) Rhythm-responsive bracelet; (b) Smart lacquer ring.}
\vspace{-2mm}
\label{fig:cowork3}
\Description{Digital lacquerware product. (a) Rhythm-responsive bracelet; (b) Smart lacquer ring.}
\end{figure*}

\begin{figure*}[htbp]
  \includegraphics[width=\textwidth]{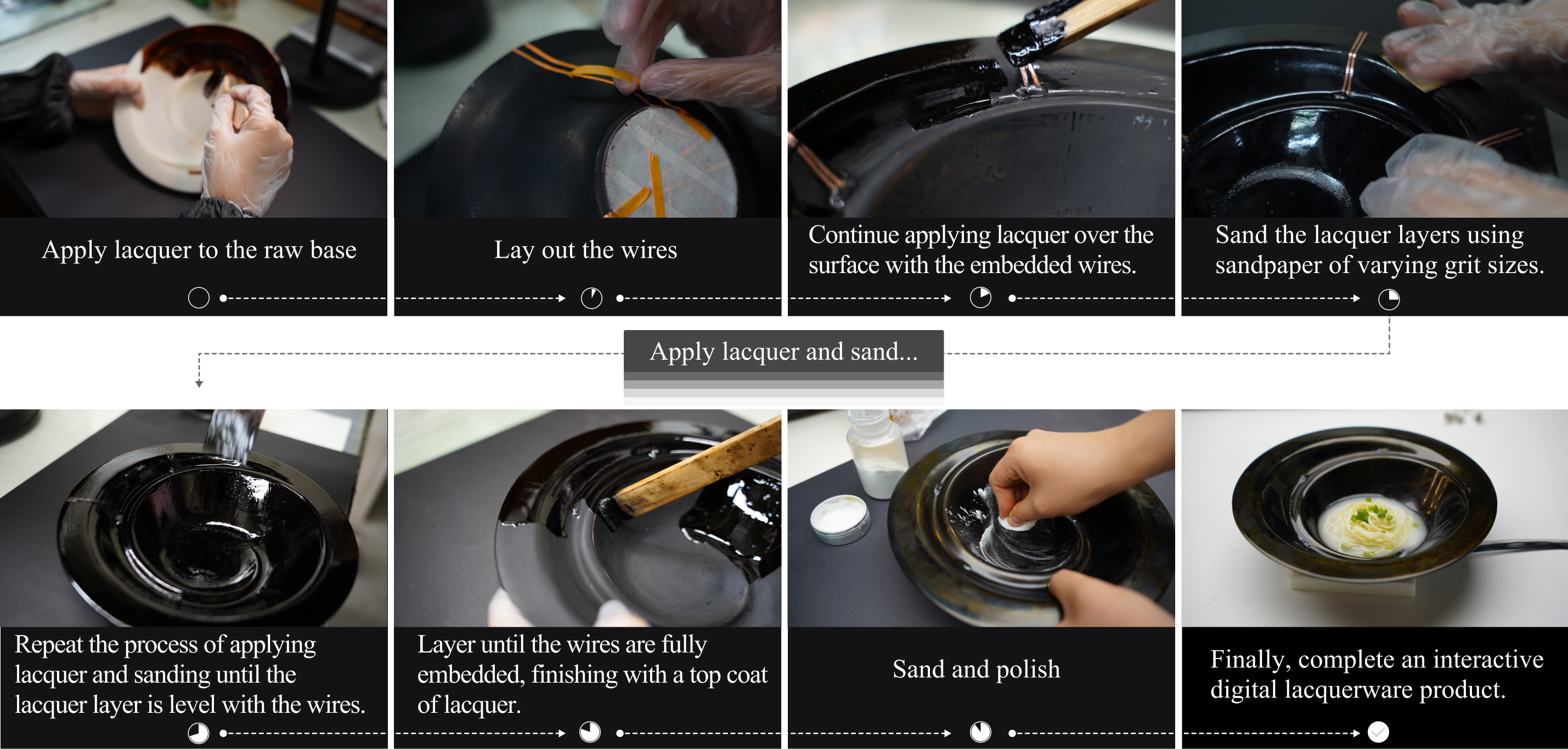}
  \vspace{-3.8mm}
  \caption{Co-creation of lacquerware artwork.}
  \vspace{-2mm}
  \label{fig:cowork process}
  \Description{Co-creation of lacquerware artwork.}
\end{figure*}

\subsubsection{Rhythm-responsive Bracelet}
A4, B4\&B9 innovatively applied the lacquer interface to a bracelet (Figure \ref{fig:cowork3}-a). Two circular copper plates inside the bracelet serve as interactive touchpoints (Table \ref{group}-d). As the dancer moves, the wrist makes contact with different touchpoints inside the bracelet, triggering various musical notes.
A4 considered the integration of copper touchpoints with the bracelet to be highly ingenious. A4 also stated, "I was able to achieve my vision effortlessly, and the toolkit did not impose much technical burden beyond the traditional lacquer craft process of making the bracelet." This design enhances dancers' expressive range through the fusion of art and technology, offering a more immersive and sensorially rich viewing experience.

\subsubsection{Smart Lacquer Ring}
A5\&B5 co-designed a smart ring (Figure \ref{fig:cowork3}-b). Sliding the thumb over the three copper plates adjusts the volume, while pressing the central circle starts or stops playback (Table \ref{group}-e). "This is a pattern design with ethnic flair," remarked A5. Due to the small size of the ring, B5 carefully designed the arrangement of the wires and consulted with A5 to determine the optimal placement of the copper sheet, balancing aesthetics with the most suitable position for the wearer’s finger.

"Wearable devices are not merely functional items; they are also emblems of personal style and identity," stated A5. The unique cultural and aesthetic values of lacquer art also highlight its innovative potential in wearable devices. Lacquer’s versatility and flexibility offer high customization potential for wearable devices, allowing users to tailor the design to their aesthetic preferences. This contrasts with the standardized designs of mass-produced digital devices, providing a more unique and personalized experience for users.

\subsubsection{Braille Tactile Interface}
Drawing from the unique tactile qualities of lacquer, A6\&B6 designed a Braille Tactile Interface (Figure \ref{fig:cowork2}-c). With B6's assistance, sensors were embedded within the lacquer layer (Table \ref{group}-f). Then A6 sculpted the Braille word "PRESS" and put over the sensors, then coated with multiple layers of lacquer and polished to form a smooth lacquer interface. Wires were affixed to the underlying lacquer surface and routed to the back of the board. By touching and pressing the textured areas on the surface, visually impaired users can naturally operate the light switch through tactile interaction.

Visually impaired individuals rely on their sense of touch to perceive the world, capable of more precisely discerning the tactile feedback of different materials \cite{10.1145/3613905.3650960}. However, existing Braille interface designs often consist of cold, hard materials. The Braille interface crafted from lacquer not only enriches the tactile experience for visually impaired person but also enhances the equal interactive experience between blind and sighted individuals through the warm texture of lacquer, promoting the potential of digital crafts in inclusive design directions.

\subsection{Result}

\subsubsection{Toolkit Usability Evaluation}
We collected questionnaires from 7 artisans and 9 HCI researchers, evaluating the toolkits and the user experience in usability, practicality, and cross-disciplinary collaboration. Both groups highly rated the toolkits, with HCI researchers scoring marginally higher than artisans.

\begin{figure*}[htbp]
\centering
\includegraphics[width=\textwidth]{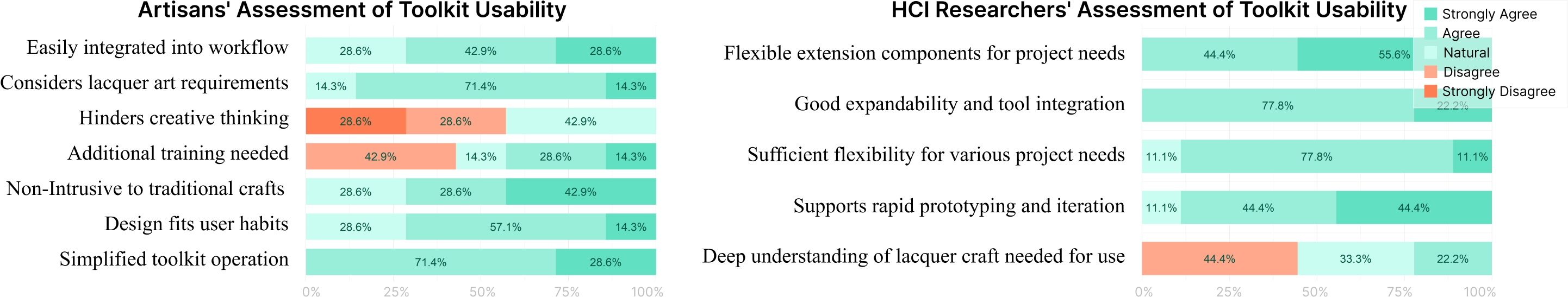}
\vspace{-3mm}
\caption{Evaluation from usability.}
\label{fig:zzp1}
\Description{Evaluation from usability.}
\end{figure*}

Regarding usability, the average score from artisans was 3.65 (SD = 0.76), while the HCI researchers gave an average score of 3.98 (SD = 0.70). Artisans appreciated the simplicity of the toolkit, stating that the steps are clear and straightforward (t = 6.97, p < 0.05) and that using the toolkit did not disrupt their creative flow (t = -2.52, p < 0.05). HCI researchers emphasized that the toolkit supported rapid prototyping and iteration (t = 5.66, p < 0.05). This highlights the fact that artisans focused on the toolkit’s operational ease and creative support, while HCI researchers prioritized efficiency in design and technical implementation (Figure \ref{fig:zzp1}).

\begin{figure*}[htbp]
\centering
\includegraphics[width=\textwidth]{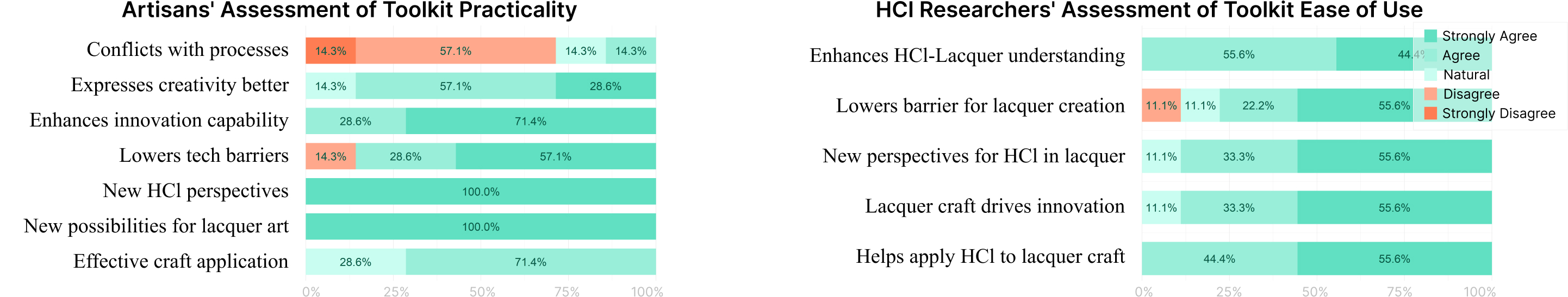}
\vspace{-3mm}
\caption{Evaluation from practicality.}
\label{fig:zzp2}
\Description{Evaluation from practicality.}
\end{figure*}

Regarding practicality, the average score from artisans was 4.18 (SD = 0.88), while HCI researchers a higher score of 4.42 (SD = 0.11). HCI researchers universally recognized the practicality of the toolkits, while artisans gave the highest scores (5.0) for "opening new possibilities for innovation" and "offering new perspectives or methods." HCI researchers agreed that the toolkit effectively applied HCI interaction principles (t = 8.85, p < 0.05) and enhanced their understanding and ability to innovate in the fusion of traditional crafts with interactive technology (t = 8.22, p < 0.05). Both groups consistently praised the toolkit’s practicality (Figure \ref{fig:zzp2}).

\begin{figure*}[htbp]
\centering
\includegraphics[width=\textwidth]{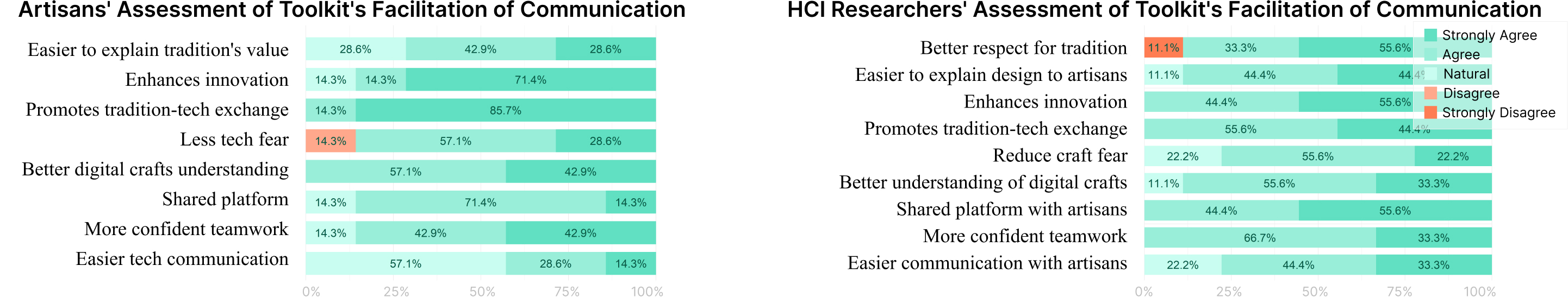}
\vspace{-3mm}
\caption{Evaluation from interdisciplinary communication and collaboration.}
\vspace{-2mm}
\label{fig:zzp3}
\Description{Evaluation from interdisciplinary communication and collaboration.}
\end{figure*}

Regarding facilitating communication, the average score from artisans was 4.21 (SD = 0.38), while HCI researchers rated it at 4.31 (SD = 0.19). Both groups agreed that the toolkit facilitated communication and enabled quick collaborative outputs. Artisans felt that the toolkit increased their confidence in collaborating with cross-disciplinary teams (t = 4.50, p < 0.05), and HCI researchers highlighted that the toolkit significantly boosted innovation in cross-disciplinary projects and enhanced their understanding of lacquer art (t = 5.50, p < 0.05), effectively enhancing efficient communication and collaboration (Figure \ref{fig:zzp3}).

During the interviews, participants generally regarded the technical principles as easy to understand. A1, A2, B1 and B2 compared their two version toolkit experiences and noted the improvement significantly reduced cognitive load. Artists stated that the toolkit didn't disrupt their lacquer application or pattern design. While adding circuits needed additional consideration, wire layers could be concealed well and touchpoints could be artistically enhanced by design (A3, A4), being less of a constraint and even becoming part of the creative process (A3). B6 highlighted that certain sensors, such as pressure sensors, could detect capacitance changes through deformation without exposing touchpoints, offering more creative possibilities. The artist group found that current interactive functionalities met their basic needs but suggested adding diverse examples to showcase potential interaction effects, helping them explore more technical options. They also expressed interest in an online consultation service with digital professionals for further support (A3, A4, A5, A7, B5, B9).

Participants highlighted two main concerns: circuitry concealment and stability. For concealment, artists sought seamless integration of circuits to avoid disrupting aesthetics. One artist remarked, "Sometimes we use transparent lacquer for the surface, and in such cases, the wiring must be handled very carefully" (A2). For stability, there were concerns about circuit reliability. An artist noted, "If the circuit gets damaged during air-drying or sanding, it could disrupt the entire production cycle" (B3). Participants also stressed the importance of repairability, with one asking, "If something goes wrong, can the circuit be repaired, and at what cost?" (A5).

For larger works, participants suggested allowing some circuits to be exposed, but emphasized the need for customized toolkits tailored for large-scale projects (A1, A2, A4, A5, A7, B4, B9). HCI researchers noted that connecting interactive circuits often requires iterative adjustments, so the system design should support multiple iterations (B5). This aligns with our earlier findings. Given the high cost of lacquerware production, participants recommended developing a habit of timely testing during the creation process (A2, A3, A6, B5, B8).

\subsubsection{Bidirectional Insights from Craft-Tech Alignment Manual}

%

Layered Interactions fostered bidirectional inspiration between artists and HCI resrerchers. While the creative process largely followed traditional methods, the introduction of interactive functions reshaped how artists conceptualized effects. A1 noted, “Interactive technology forces us to consider the integration of form, materials, and interactive effects at the early design stage, making the process more comprehensive and engaging.” A2\&A5 also explored how lacquer art could be integrated into everyday interactive scenarios, such as in the designs for the Braille interface and lacquer rings. These collaborations demonstrated the potential to move beyond traditional lacquerware functionality and integrate features relevant to modern life (A5, A6). This aligns with the toolkit’s original intent of enabling non-destructive techniques.
Conversely, the technical challenges posed by artists inspired new perspectives for engineers. B7 commented, “Artists brought a fresh viewpoint to circuit design. Their creativity not only redefined how we think about circuit functions but also introduced a new aesthetic dimension to our designs”.

The toolkit simplified the mapping between digital technologies and craft processes by decomposing workflows into detailed steps and facilitating efficient technical solutions. This clarity helped define the roles of artisans and HCI researchers, fostering mutual understanding. A3 remarked, “The modular design improved communication and collaboration, allowing me to identify the technical solutions needed to realize my ideas while helping technical partners understand the limitations and needs of the craft”. B5 added, “The toolkit helped us quickly establish a shared understanding from concept to practice, identifying feasible combinations of techniques and crafts... Emphasizing that the blend of creativity and technology is equally important in the craft domain”.

Modular design supported rapid prototyping, enabling creators to quickly identify areas with greater creative potential, thereby improving efficiency. Adjustments focused on effect selection and parameter tuning, which both artists and technical personnel found manageable. They expressed a willingness to continue exploring and adopting the technology (A1, A4, A5, B2, B7, B8, C3). A5 noted, “The modular tools let us test different steps quickly without worrying about wasting time on repetition. This allowed me to focus on creativity rather than being constrained by technical workflows”. B2 echoed this sentiment, stating, “The system maintains relatively independent working spaces but allows timely communication and adjustments in overlapping areas. This dynamic flexibility fulfilled my long-standing desire for more adaptable collaboration”.

\subsubsection{Non-Intrusive Methods Aligned with Craft Processes}
The craft-based integration approach was widely recognized by both artisan artists and HCI researchers, prompting reflections on digital craftsmanship. Participants noted that the fusion of craft and technology goes beyond the artifact, fostering engagement between creators and audiences through real-time feedback for a more interactive and enjoyable experience (A3, A4, B3, C3). Regarding "non-intrusive" integration, most participants agreed it involves preserving the core artistic and functional characteristics of traditional crafts. Traditional artists prioritized seamless fusion, while avant-garde artists were more open to minor disruptive changes. This acceptance also correlated with age, as older artisans were often less receptive to change (A1, A4, A6).

Participants noted multiple dimensions of non-intrusive integration. As B7 remarked, “The entry point is crucial. This experiment showed me how to seek holistic integration of ideas, materials, processes, and technologies”. Another commented, “The key is to use technology’s value while highlighting the uniqueness of lacquer art, avoiding integration for its own sake” (A6).

The visibility of exposed touchpoints in Layered Interactions was seen by artists as an opportunity rather than a limitation. A3 noted, “With aesthetic design, exposed touchpoints can become highlights of the work.” They also suggested defining technological possibilities and limitations early in the creation process, such as reserving touchpoints or designing the base structure, to optimize workflows. “Understanding technical constraints early allows us to turn them into creative inspiration” (A2, B1, B3, B9). Another noted, “The inspiration brought by digitization may impact artisans more profoundly than expected by digital practitioners” (A3, A4, A5, A6, B2). A5 remarked, “Digitization has made changes to static lacquer art that once seemed impossible.” A3 further emphasized, “New ideas and techniques expand lacquer art’s expressive potential.”

For technical personnel, this work enhanced the realism of digital technologies. As B3 noted, “Traditional digital craftsmanship often relies on predefined datasets like 3D modeling or AI training, but physical artifacts bring stronger ties to creativity, maintaining a close connection between craftsmanship and humans.” This realism offers HCI experts new tactile media for expression, even introducing the possibility of "touchable AI".

HCI researchers suggested that GUI interfaces could help artisans adjust effects more intuitively (B1, B4, B5). They also recommended exploring node-based or blueprint-style programming approaches, similar to TouchDesigner (B3), and proposed developing AI agents to guide code modification as a promising direction (B5).

\section{Discussion}

\subsection{Exploring the Unique Design Space of Lacquer Art}
This paper delves into the under-researched domain of lacquer art in the digital craft landscape, aiming to drive interactive design through the intrinsic characteristics of the craft while achieving a seamless integration of traditional craftsmanship with modern technology. A pivotal question in digital crafts is how to integrate technology with the unique attributes of craftsmanship organically. Compared to previous work, we propose a non-intrusive perspective, and introduces three key dimensions for digital craft design guidelines:
\begin{enumerate}
    \item {\emph{Material Non-Intrusiveness}}: Balancing technical and craft constraints is essential. The requirements of lacquer art (e.g., lacquer thickness, coating uniformity) must be considered alongside those of circuit design (e.g., conductivity, sensor sensitivity). Through experimentation and adjustments, we ensure circuit functionality while maintaining the artistic effect. For instance, thin circuit components (e.g., flexible circuit boards, thin sensors) are chosen to minimize their impact on lacquer thickness, and circuit layouts are concealed to integrate the technology within the lacquer layers, especially in decorated lacquer areas.
    \item {\emph{Craft Process Non-Intrusiveness}}: Technology should be seamlessly embedded into the traditional lacquer crafting process. For example, circuit installation can be integrated into the brushing steps without disrupting the artisan's workflow. At the same time, it is necessary to measure the circuit according to the process steps to ensure that artisans can adjust the lacquer layers as needed in different creative scenarios.
    \item {\emph{Digital Integration Posture Non-Intrusiveness}}: This represents an ethical consideration, positioning technology as an auxiliary rather than a replacement tool, respecting the habits of artisans and offering more user-centric care. The artisan remains central to the process, and the toolkit supports this by offering defined modification areas that maintain overall consistency while preserving artistic freedom.
\end{enumerate}

Understanding the craft processes and artisans has prompted HCI practitioners to reflect on and re-evaluate the importance of preserving these processes. Firstly, maintaining these processes can alleviate tensions between traditional artisans and technology. Many digital craft studies treat digital tools as rapid replacements or improvements to manual labor, emphasizing efficiency and precision gains while overlooking artisan habits. This approach may create a gap and preven the effective integration of digital tools into the production processes of the craft industry. Adhering to craft processes can lead to better acceptance by artisans, easing technological tensions and revitalizing crafts. 
Secondly, traditional craft workflows offer a robust framework for digital craft design, preserving the artistry and cultural heritage embedded within. These processes and techniques, refined over centuries, maximize material properties and aesthetics. Furthermore, starting from the framework of traditional crafts, it sustains the artist's control over the artwork, enabling them to both mentally envision and physically interact with the work's state, such as sensing lacquer layer thickness and temperature through touch, or observing its drying process. Additionally, employing HCI research methods and standards makes the tacit knowledge of traditional crafts more comprehensible. By standardizing craft and technology, we facilitate a better alignment between traditional crafts and modern technology, aiding in the establishment of a "craft archive" for digital crafts that can be repeatedly practiced and refined, ensuring the longevity of these artisanal practices.

\subsection{Broadening HCI through Digital Craft Integration}

Traditional crafts are typically static, one-time physical creations, where mainly as as observers or owners with limited participation. This study explores methods to infus traditional crafts with interactive properties, transforming them into mediums for emotional connection. The integration of digital craftsmanship not only extends technological capabilities but also inspires new ideas and possibilities for artisans, enriching the expressive potential of their work. This concept extends beyond lacquerware to other traditional crafts and materials, offering a valuable design framework for the HCI and TUI communities. Our findings indicate strong acceptance of the toolkit among both artisans and HCI researchers, with its interaction-driven creativity inspiring new applications and audiences. Artisans proposed innovative uses, such as tableware, furniture, toys, customized wheelchair handles, clothing, and jewelry. Some participants suggested collaborating with museums and cultural institutions to explore new ways of popularizing and educating about lacquerware.

The study also broadens material options for TUI by leveraging the tactile and sensory qualities of lacquer. Lacquer evoke profound emotional responses, making them ideal for interaction design. By combining the sensitivity of traditional materials with digital technology, we expand interaction modes beyond screens and buttons to create more engaging and emotionally resonant experiences. Rooted in real-life needs, traditional crafts have developed intuitive applications over centuries. Integrating these materials into digital interactions aligns ubiquitous computing with users’ sensory and physical needs, creating more relatable interaction scenarios.

The fusion of crafts and technology prompts HCI practitioners to reflect on the relationship between technology and humanity. Craft fundamentally emphasizes the harmony between humans, materials, techniques, and nature. Integrating digital technology challenges the utilitarian focus of technology, embedding artistry, cultural significance, and emotional depth into HCI design. This interdisciplinary approach encourages HCI practitioners to view technology not only as a problem-solving tool but also as a medium for emotional, cultural, and artistic expression. As HCI designers pursue innovation, they must consider these implications, ensuring that technology supports the value and heritage of traditional crafts while maintaining sustainability.

\subsection{Cross-Disciplinary Collaboration Enabled by Modular Toolkits}
Resolving tensions between artisans and technicians is crucial in collaborative work. Artisans often focus on realizing their concept treating technology, materials, and techniques as to achieve unique, personalized artistic styles. In contrast, HCI researchers prioritize universal solutions, emphasizing innovation, scalability, robustness, and precision of tools. In interdisciplinary collaboration, differences in design objectives, needs, and values often lead to tension. Our research proposes strategies to alleviate these conflicts and foster cooperation.

Facilitating dialogue between craft and technology. It is unrealistic to expect artisans to systematically learn technical knowledge, just as it is challenging for researchers to master advanced craft techniques. A collaborative model that follows the craft process helps build mutual trust and equality between HCI researchers and artisans, improving cross-disciplinary collaboration efficiency. Artisans gain an understanding of how interactive lacquer art is achieved, while HCI researchers gain a deeper insight into the details of traditional craftsmanship, thus achieving cognitive alignment. Technicians are better able to understand the artisans' needs and workflow, enabling them to design more practical tools and technical solutions, such as the requirement for thin, stable circuit embedding to ensure long-term lacquer durability (A3, A5). Without understanding the lacquer production process, one might overlook its irreversible nature.

Clearly defines the boundaries between the work of artisans and HCI researchers. In our study, the collaboration between artisans and HCI researchers initiates with a joint discussion of project objectives, then followed by a 'relay-style' workflow." The provided craft process guidelines serve as excellent guidance, allowing collaborators to clearly define their roles and focus on their areas of expertise. Positive interactions were demonstrated in the results. In the collaboration, artisans focused on how to preserve the core of lacquer (such as its functionality, material properties) and its aesthetic qualities in digital works, while those with technical backgrounds focused on the rationality of the interactions and the optimal layout of circuits.

Providing significant flexibility to support the diversity and uniqueness of traditional craft. For example, it allows artisans to design touchpoints, whether in crescent shapes (pre-experiment) or abstract geometric forms (Group 3). By following the basic technical structure and indicating where modifications can be made,  it reduces the technical learning curve while preserving creative freedom. Our toolkit provides basic outputs (sound and light), but two sets of experiments added visuals (dynamic scenes on the screen) based on music. Furthermore, A1 (skilled in both lacquer techniques and glass making) believes that the \emph{"Craft-Tech Alignment Manual"} we offer is not limited to lacquer art but can be extended to fields like glass and woodworking, helping to bridge the gap between digital technology and craftsmanship.

\section{Limitation and Future Work}
\subsection{Limitation}
Our research faces three main limitations. First, the specialized nature of the craft limited the diversity of participants. Due to the high skill level required for lacquerwork, the number of participants was restricted. In the future, we plan to include artists from other craft backgrounds to broaden participation, increasing the diversity of data and enhancing the study's representativeness. Second, the toolkit's functionality needs further expansion. While temperature, and pressure sensing were well-received, participants expressed interest in adding distance sensor, tilt sensor, and other types of sensors. We aim to enhance the toolkit’s capabilities to accommodate more complex creative scenarios. Lastly, the toolkit’s current design is limited to small-scale works. For larger-scale projects like home decor or exhibitions, it will require customization. We also plan to explore scalability for industrial applications, as participants showed interest in mass production(A7, A2).

By addressing these improvements, we aim to transition the toolkit from small-scale lab use to large-scale real-world applications, meeting artists’ creative needs and expanding its industrial potential.

\subsection{Future Work}
In the future, we plan to develop an intuitive GUI for the toolkit, allowing artisans to easily adjust parameters like brightness, volume, and tactile sensitivity without modifying code. This will simplify operation and enhance control over their creations. We are also considering visual programming models, similar to TouchDesigner's node-based systems, lowering technical barriers to creativity.

A4 highlighted the difficulty of finding HCI collaborators with whom effective collaboration can be easily achieved. As AI technology advances, we plan to explore using large language models (LLMs) to support creative processes. By capturing the knowledge of artisans and HCI researchers, AI can suggest tools and generate creative solutions to enhance efficiency. A multi-agent system is also a potential approach, aimed at summarizing artisans' cognitive patterns and generating personalized instructions for tool usage.

Lastly, we will explore how digital tools can optimize the teaching and dissemination of lacquer craftsmanship, making it easier for more people to learn and preserve these traditional techniques.

\section{Conclusion}
This paper explores the synergy between digital technology and traditional lacquer craftsmanship. We introduces \emph{Layered Interactions}, a \emph{Non-Intrusive} digital craft method that enhances rather than replaces traditional techniques. We design a toolkit fosters interdisciplinary collaboration. Overall, the integration of digital elements increases the expressiveness, utility, and adaptability of traditional crafts in modern contexts. Additionally, leveraging the aesthetic qualities and application scenarios of lacquer art enhances the authenticity and materiality of artificial objects.

\begin{acks}
We sincerely appreciate the support of the Key Laboratory of Traditional Crafts and Materials Research in Cultural Tourism (Lacquer Craft and Materials Research), Ministry of Culture and Tourism, Tsinghua University. This project is supported by the "Dual High" Project of Tsinghua Humanity Development (No.2023TSG08103).
\end{acks}

\bibliographystyle{ACM-Reference-Format}
\bibliography{reference}

\end{document}